\documentclass[multphys,vecphys]{svmult}

\newcommand{\Index}[1]{\index{#1}#1}

\usepackage{makeidx}     
\usepackage{graphicx,psfig}    
\usepackage{multicol}    
\usepackage{amsmath}
\usepackage{latexsym}
\usepackage{float}
\usepackage{amssymb}
\usepackage[latin1]{inputenc}
\usepackage[T1]{fontenc}

\makeindex             

\begin{document}

\title*{THEORIES OF THE STRUCTURAL GLASS TRANSITION}
\author{Rolf Schilling}
\institute{Institut f\"ur Physik, Johannes Gutenberg-Universit\"at
Mainz, Staudinger Weg 7, D-55099 Mainz, Germany \\
\texttt{Rolf.Schilling@uni-mainz.de} }
%
%
\maketitle


\section{Introduction}
Equilibrium phase transitions, e.g. the transition at 0$^{\circ}$C
from water to an ice-crystal, are common phenomena in nature. Such
phase transitions between a disordered high temperature phase and
an ordered low temperature one are rather well understood and can
be theoretically described within statistical mechanics. Given the
interaction between the species (particles, spins, etc.) the
partition function can be calculated, in principle. Its logarithm
yields, e.g. for a canonical ensemble, the free energy which is
singular at the equilibrium transition point. This allows to fix
this point from first principles.

Besides such \Index{order-disorder transitions} there also exist
transitions between disordered phases. Excluding liquid-liquid
transitions these will be called \textit{glass transitions}. A
prominent example is the transition from a supercooled melt of
SiO$_2$-molecules to an amorphous phase which is the well-known
``window glass''. One distinguishes two types of glass
transitions: \Index{spin glass transitions} \cite{1rs,2rs} and
\Index{structural glass transitions} \cite{3rs,4rs,5rs,6rs}.
Their main difference is that the former occur mostly in systems
with \textit{quenched disorder} and for the latter the disorder is
\textit{self-generated}. We stress that this classification should
not be taken too strict since there are also models
\textit{without} quenched disorder showing spin glass behavior
\cite{7rs}. Typical systems undergoing a structural glass
transition are liquids, particularly molecular liquids, like the
famous example of SiO$_2$. In recent years investigations of
specific \Index{spin glass models}, e.g. \Index{Potts glass} and
so-called \index{p-spin models}$p$-spin models
with $p \geq 3$ where $p$ spins
are coupled by randomly frozen-in (i.e. quenched) \textit
{infinite range interactions}, have revealed some similarities
with structural glasses \cite{8rs}. We will come back to this
point in the 4th chapter.

In contrast to conventional order-disorder transitions, glass
transitions are less well understood. There is a broad consensus
that a spin glass transition exists and that an appropriately
disorder-averaged free energy is singular at the transition point
in case of mean field models, i.e. models with infinite range
interactions or infinite dimensions. But this is less obvious for
short range interactions. The situation for the structural glass
transition is even less satisfactory, although substantial
progress has been made in the last two decades.

In this article we will mainly focus on the structural glass
transition. Spin glass behavior and spin glass transition will be
discussed in this monograph by H. Horner. For further details and
references on spin glasses, the reader may consult his
contribution.

As mentioned above, structural glasses can be obtained by cooling
a liquid. In order to bypass crystallization one has to choose a
\textit{finite} cooling rate. For good glassformers, like SiO$_2$,
this rate can be rather modest whereas bad glassformers, like most
metallic glasses, require extremely high cooling rates. Under such
a cooling process the shear viscosity $\eta (T)$ increases. Close
to the so-called 
\index{calorimetric glass transition}{\it calorimetric} glass transition
temperature $T_g$ the supercooled liquid falls out of equilibrium
and becomes a glass. At $T_g$ thermodynamical quantities like
density $n(T)$, specific heat $c_p(T)$ at constant pressure,
etc.~show cross-over behavior, i.e. the slope of $n(T)$ and
$c_p(T)$ make a more or less well pronounced jump, depending on
the cooling rate. $T_g$ itself depends on the cooling rate, too.
Although $T_g$ plays an important practical role, it is less
interesting from a fundamental point of view, due to its cooling
rate dependence. Besides $T_g$ there are at least three more
characteristic temperatures $T_0,T_K$ and $T_c$. For many glass
formers, $\eta (T)$ can be fitted by the
 \Index{Vogel-Fulcher-Tammann law}

\begin{equation}\label{eq1rs}
\eta (T) = C \exp \frac {A}{k_B(T-T_0)} \quad , \; T \geq T_0
\end{equation}

\noindent with $A > 0$ and $C > 0$. The shear viscosity diverges
at $T_0$. Extrapolating the \Index{excess entropy}
$S_\textrm{excess}(T)$ of the supercooled liquid with respect to
the crystalline phase to lower temperatures there is the so-called
\Index{Kauzmann temperature} $T_K$ at which $S_{\textrm
{excess}}$ vanishes:
\begin{equation}\label{eq2rs}
S_{\textrm{excess}}(T_K)=0 \quad .
\end{equation}

\noindent Since it is argued that a disordered phase should not
have a smaller entropy than the crystalline one,
$S_\textrm{excess}$ can not become negative. Therefore, the system
has to undergo a \textit {static} glass transition at $T_K$. This
conclusion, however, is not compelling, since there exists inverse
melting, i.e.~liquids freeze when heated or crystals melt when
cooled \cite{9rs}. In that case the total entropy of the crystal
is higher than that of the liquid. A recent discussion of the
Kauzmann problem can be found in Ref. \cite{10rs}. $T_g, T_0$ and
$T_K$ have played an essential role for many decades. In 1984
quite a new theoretical approach, the \Index{mode coupling theory}
\cite{11rs}, has shown that there is a critical temperature $T_c$,
at which a \textit {dynamical} glass transition takes place. One
of the main features is that the
\index{nonergodicity parameter}nonergodicity parameters 
$f(\vec{q},T)$ which can be considered
as glass order parameters change discontinuously at $T_c$:

\begin{equation} \label{eq3rs}
f(\vec{q},T) = \left \{ \begin{array} {r@{\quad , \quad}l} 0 &
T>T_c \\ >0 & T\leq T_c \quad .
\end{array} \right.
\end{equation}
Since then numerous experimental investigations and computer
simulations were stimulated (see reviews \cite{12rs,13rs} and Ref.
\cite{14rs,15rs}). They have shown new characteristic \textit
{dynamical features} close to the dynamical glass transition point
$T_c$ consistent with mode coupling theory (see also the
contribution by U. Buchenau in this monograph).

This short exposition of some of the characteristics of glassy
behavior should have given a first impression on how diverse the
phenomena in the glass transition region can be. Therefore it is
obvious that a successful theoretical description which covers all
facets is extremely hard. There are mainly two possible
theoretical approaches: phenomenological or microscopic ones.
\textit {Phenomenological theories} start from some of the
phenomena of glasses, and are named thereafter. Based on these
phenomena a theoretical description is developed capable of
describing the observed phenomena. In several cases appealing
``physical pictures'' are used. However, a couple of assumptions
are made which are not proven. The predictive power of such
phenomenological approaches is rather limited. This is quite
different from a \textit{microscopic theory}. By microscopic we
mean that the physical quantities can be calculated from first
principles if the interactions between the species are given.
Since the glass transition region is located at rather high
temperature quantum effects can be neglected. Therefore, a
microscopic theory starts from a classical $N$-body problem. The
next chapter will discuss some of the phenomenological theories.
The major part of this article is devoted to microscopic theories:
The 3rd chapter describes mode coupling theory and the 4th chapter
the replica theory for structural glasses.

Finally we want to stress that the present contribution presents a
selection and does not aim to be complete. This holds mainly for
the phenomenological models. We also do not discuss the
\Index{potential energy landscape}``potential energy landscape'' 
approach \cite{16rs} which
recently has led to new interesting results \cite{17rs,18rs}.
Almost all of them were obtained from computer simulations. It
would be desirable to complement these investigations by
analytical theories.

\section{PHENOMENOLOGICAL APPROACHES}
The presentation in this chapter will be rather short. More
details and additional phenomenological approaches can be found in
the monographs \cite{3rs,5rs,6rs} and in the review \cite{4rs}.

\subsection{Adam-Gibbs Theory}
In 1965 Adam and Gibbs suggested a theory based on the assumption
that \Index{dynamically cooperative regions} occur when
decreasing the temperature towards the glass transition point
\cite{19rs}. The particles in these regions perform cooperative
motion, which leads to a reduction of the configurational degrees
of freedom. Here we follow the description of \Index{Adam-Gibbs
theory} as given in Ref.~\cite{4rs}. The basis of that theory is a
number of assumptions:

\begin{itemize}
\item[{1)}] For given temperature $T$ there are dynamically
cooperative regions labeled by $1,\ldots,i,\ldots, n(T)$ with
$N_1(T),\ldots,N_i(T),\ldots,N_{n(T)}(T)$ particles. The number
$n(T)$ of regions decreases with decreasing temperature (see
Fig.~\ref{fig1rs}). Due to the conservation of $N$, the total
number of particles, it is:

\begin{equation}\label{eq4rs}
\sum \limits _{i=1}^{n(T)} N_i(T)=N  \quad .
\end{equation}

\item[{2)}]
These regions take two configurations, only. Accordingly the
entropy $s$ per region is:

\begin{equation}\label{eq5rs}\
s=k_B \ln 2 \quad .
\end{equation}

The assumptions that each region takes two configurations, only,
is not crucial, but it should be a \textit {finite} number of
order one.
\item[{3)}]
Fluctuations of $N_i(T)$ are small, i.e. it is:
\begin{equation}\label{eq6rs}
N_i(T) \approx N_0(T)=N/n(T) \quad .
\end{equation}

because of Eq.~(\ref{eq4rs}).

With these three assumptions we can relate the average number $N_0
(T)$ of particles within a dynamically cooperative region to the
\Index{configurational entropy}:

\begin{equation}\label{eq7rs}
S_c (T)=k_B \ln \, [{\textrm{number of total configurations}}]
\quad .
\end{equation}

\begin{figure}[t]
\unitlength1cm
\begin{picture}(11,6)
\centerline{\psfig{file=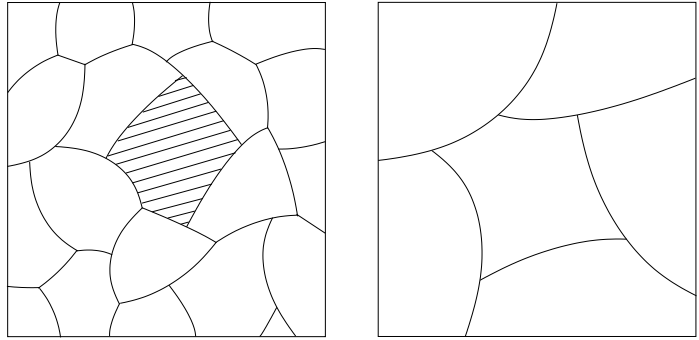,angle=-90,width=10cm}}
\end{picture}
\caption{ \label{fig1rs} Illustration of dynamically cooperative
regions (one of them is shown as hatched). Left part: high
temperature; right part: low temperatures }
\end{figure}

With $2^{n(T)}$, the number of total configurations we get from
Eq.~(\ref{eq7rs}) by use of Eq.~(\ref{eq6rs}):

\begin{equation}\label{eq8rs}
N_0(T)= N k_B \ln 2 / S_c(T)
\end{equation}

i.e.~$N_0(T)$ and therefore the size of those regions is inversely
proportional to the configurational entropy. This is plausible
since the number of possible configurations decreases with
increasing $N_0(T)$.

The next important steps are to assume:

\item[{4)}]
There is a temperature $T_K$ at which $N_0(T)$ becomes infinite
(for $N = \infty$), Then Eq.~(\ref{eq8rs}) implies

\begin{eqnarray} \label{eq9rs}
S_c (T)= \left \{\begin{array}{ll} > 0  \quad ,  \quad T>T_K\\
=0 \quad , \quad T \leq T_K \quad , \\
\end{array} \right.
\end{eqnarray}

i.e~ $T_K$ is the \Index{Kauzmann temperature} \cite{20rs}.

\item[{5)}] The transition between both configurations of a region
is an activated process with an activation energy:

\end{itemize}
\begin{equation}\label{eq10rs}
E(T)= e_0N_0 (T)
\end{equation}

\noindent where $e_0$ may be weakly $T$-dependent. Then the
transition time $\tau(T)$ is given by:

\begin{equation} \label{eq11rs}
\tau(T) \sim \exp (\beta E (T)) = \exp (\beta e_0 N_0(T)) \quad .
\end{equation}

\noindent Substituting $N_0$ from Eq.~(\ref{eq8rs}), we arrive at:

\begin{equation} \label{eq12rs}
\tau(T) \sim \exp \left[\frac{e_0 N  \ln 2} {T S_c (T)}\right] .
\end{equation}

\noindent Assuming that $S(T)$ vanishes linearly at $T_K$ we
obtain from Eq.~(\ref{eq12rs}) the \Index{Vogel-Fulcher-Tammann
law} close to $T_K$:

\begin{equation} \label{eq13rs}
\tau(T) \sim \exp \left[\frac{e_0 N \ln 2} {T_K S'_c (T_K)
(T-T_K)}\right]
\end{equation}

\noindent with $T_0=T_K$.

\noindent This example demonstrates that the appealing picture of
dynamically cooperative regions in combination with several
assumptions allows to derive the Vogel-Fulcher-Tammann law. But it
also shows that no additional predictions are made. Furthermore,
it is obvious that the growth of the dynamically cooperative
regions would be accomplished by a divergent length scale, which,
however, has never been found in experiments or computer
simulations.


\begin{figure}[t]
\unitlength1cm
\begin{picture}(11,12)
\centerline{\psfig{file=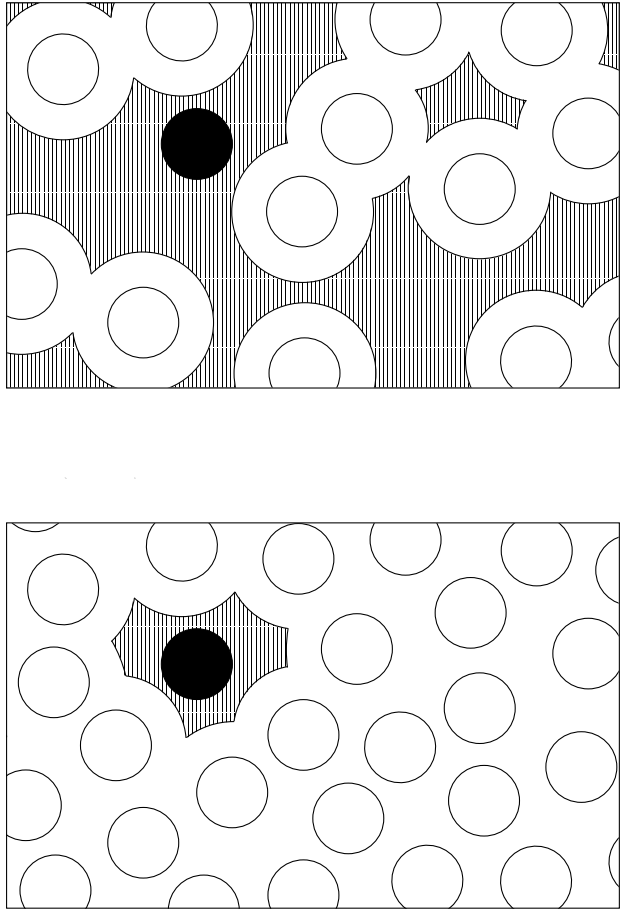,angle=0,width=9cm}}
\end{picture}
\caption{ \label{fig2rs} Illustration of the free volume of one
sphere (full circle) within a fixed configurations of other
spheres (open circles). The free volume is shown as hatched
regions. Upper panel: $T>T_0$ for which the free volume forms a
percolation cluster. Lower panel: $T<T_0$ for which the free
volume does not percolate but gets localized. }
\end{figure}

\subsection{Free-volume theory}
\label{sec:2}

This phenomenological description has been made by Cohen and
Turnbull in 1959 \cite{21rs}. In order to illustrate their idea we
choose a system of \textit{hard} spheres with average density $n$.
Let us fix the positions of all spheres except of the $i$-th
sphere. Then the $i$-th sphere can move freely in the so-called
\textit{free volume} $\upsilon_f(i)$ (see Fig.~\ref{fig2rs}).
Although $\upsilon_f(j)$ is correlated with $\upsilon_f(i)$  for
$i\neq j$ one assumes:

\begin{itemize}
\item[{1)}] Every sphere has a \Index{free volume} $\upsilon_f$:

\begin{equation} \label{eq14rs}
i \rightarrow  \upsilon_f(i)
\end{equation}

\item[{2)}] $\upsilon_f(i)$ are independent random numbers with
probability density  $p(\upsilon_f)$ which is assumed to be
exponential:

\begin{equation} \label{eq15rs}
p(\upsilon_f)= \mathcal{N} \exp (- \upsilon_f/ \bar{\upsilon}_f)
\end{equation}

\noindent with the mean free volume per particle:

\begin{equation} \label{eq16rs}
\bar{\upsilon}_f = \frac{1}{N} \sum\limits_{i=1}^N \upsilon_f(i)
\end{equation}

\noindent where $\mathcal{N}$ is a normalization constant.

\item[{3)}] It is obvious that $\upsilon_f(i)$ and therefore
$\bar{\upsilon}_f$ decreases with increasing density or decreasing
temperature (in case of ``soft'' particles). An essential
assumption is that there is a temperature $T_0>0$ such that the
mean free volume vanishes at $T_0$:

\begin{eqnarray} \label{eq17rs}
\bar{\upsilon}_f(T)= \left \{\begin{array}{ll} \alpha (T-T_0) \,\,\,\ ,  \,\,\,\  T \geq T_0\\
\,\,\,\ 0  \,\,\,\ \,\,\,\ \,\,\,\ \,\,\,\ \,\,\,\ , \,\,\  T < T_0\\
\end{array} \right.
\end{eqnarray}

\noindent with $\alpha$ the expansion coefficient.

\item[{4)}] The inverse shear viscosity is proportional to the probability
that the free volume per particle is larger than a certain value
$\upsilon_f^*$ , i.e.:
\end{itemize}

\begin{equation} \label{eq18rs}
\eta^{-1} (T) \sim \int\limits_{\upsilon_f^* }^{\infty} \, d
\upsilon_f p (\upsilon_f) .
\end{equation}

\noindent Substituting Eq.~(\ref{eq15rs}) and Eq.~(\ref{eq17rs})
into Eq.~(\ref{eq18rs}) we get the \Index{Vogel-Fulcher-Tammann
law}:

\begin{equation} \label{eq19rs}
\eta (T) \sim \exp \left[ \frac{\upsilon^*_f}{\alpha
(T-T_0)}\right].
\end{equation}\\

\subsection{Extended free-volume theory}

The \Index{free volume theory} has been extended by Cohen and
Grest \cite{22rs}. The main idea of this extension is to make a
connection between the glass transition and \Index{percolation}.
If the density $n$ (temperature $T$) is below $n_g$ (above $T_g$)
the individual free volumes $\upsilon_f(i)$ overlap or, in other
words, the free volume \textit{percolates} in such a way that a
(macroscopic) 
\index{percolation cluster}\textit{percolation cluster} 
exists (see
upper panel of Fig.~\ref{fig2rs}). In this case a particle can
move macroscopic distances, i.e~the diffusion constant is finite.
Now, increasing $n$ (decreasing $T$) there may exist a critical
value $n_g$ (or $T_g)$ at which the percolation cluster disappears
which implies that the particles become localized in a glass phase
with zero diffusion (see lower panel of Fig.~\ref{fig2rs}). If
this scenario would be correct the glass transition point would
coincide with the percolation threshold of the free volumes
$\upsilon_f(i)$. Since percolation is well understood \cite{23rs}
this relationship would imply several characteristic features,
e.g.~the fractal nature of the percolation cluster, close to $n_g$
(or $T_g$). However, there is no experimental evidence for such a
fractal behavior. The main critique on the free volume theory is
that the density used in the lower panel of Fig.~\ref{fig2rs} to
demonstrate the localization of the free volume still corresponds
to a liquid. Computer simulations show that the density of a
liquid of e.g.~hard discs is already higher than that used in the
lower panel of Fig.~\ref{fig2rs}. Accordingly, the free volume, as
defined above, is already localized in the liquid phase.

\subsection{Gibbs-DiMarzio theory}

This approach is not really phenomenological. But the model which
has been studied by Gibbs and DiMarzio is very special and has a
limited range of applicability. Therefore it is included in this
chapter. The model by Gibbs and DiMarzio is capable to explain the
vanishing of the configurational entropy \cite{24rs}. A system
with $N$ polymers on a cubic lattice with $N_s$ lattice sites is
considered. Each polymer consists of $l$ monomers and can form
conformations labeled by $1,2, \cdots, i, \cdots .$ Let $n_i$ be
the number of polymers with conformation $i$ and $P( \{n_i\}, N,
N_s)$ the probability that $N$ polymers on a lattice with $N_s$
sites built a set $\{n_i \}$ of conformations. Associating an
energy $\varepsilon_i$ with each individual conformation the total
energy $E$ is given by:

\begin{equation} \label{eq20rs}
E= \sum\limits_i n_i \varepsilon_i \quad .
\end{equation}

\noindent Then the number of configurations $\Omega (E, N, N_s)$
with energy $E$ is:

\begin{equation} \label{eq21rs}
\Omega (E, N, N_s) = \sum\limits_{\{n_i\}} \, P(\{n_i\}, N, N_s)
\,\, \delta (E-\sum\limits_i \varepsilon_i n_i)
\end{equation}

\noindent from which one obtains the \Index{configurational
entropy}:

\begin{equation} \label{eq22rs}
S_c (E, N, N_s) = k_B \, \ln \Omega (E, N, N_s) \quad .
\end{equation}

\noindent Of course, the nontrivial problem is to determine $P$
and to perform the sum over $\{n_i\}$ in Eq.~(\ref{eq21rs}).
Making use of a mean field approximation due to Flory and Huggins
\cite{25rs} one finally gets a critical energy $E_K$, which
corresponds to the \Index{Kauzmann temperature} $T_K$ with:

\begin{eqnarray} \label{eq23rs}
S_c (E, N, N_s)= \left \{\begin{array}{ll} > 0  \quad ,  \quad E>E_K\\
=0 \quad , \quad E \leq E_K \quad .
\end{array} \right.
\end{eqnarray}

\noindent Since mean field approximations tend to produce phase
transitions even for systems which do not show such transitions,
the validity of results Eq.~(\ref{eq23rs}) is not obvious.

\section{MICROSCOPIC THEORY: MODE COUPLING THEORY}

We hope that the short presentation in the 2nd chapter has
demonstrated that the status of those phenomenological theories is
not satisfactory. Although, plausible ``physical pictures'' are
involved they are based on a couple of crucial assumptions.
However, the validity of those remain completely unclear. This
demands for a \textit{microscopic} approach based on \textit{first
principles}. For the first time, such an approach to structural
glass transitions was made in 1984 by Bengtzelius, G\"otze and
Sj\"olander \cite{11rs}. Starting from the liquid side these
authors applied the \Index{mode coupling theory} (MCT), developed
by Kawasaki \cite{26rs} in order to describe the critical slowing
down close to a critical point, to the relaxation of the density
fluctuations in a supercooled liquid. The result is an equation of
motion (see below) for the \Index{intermediate scattering
function} $S (\vec{q}, t)$ for a \textit{simple} liquid which can
be measured by neutron- and light scattering (see the
contributions by U. Buchenau and J. B. Suck in this monograph).
The properties of the solution $S(\vec{q}, t)$ of the
MCT-equations were mainly investigated in great detail by G\"otze
and his coworkers. These results can be found in the reviews
\cite{27rs,28rs,29rs,12rs,13rs} and in their references. Since
most glass formers are molecular systems which involve
translational \textit{and} rotational degrees of freedom, MCT had
been extended to a single linear molecule in a simple liquid by
Franosch et al.~\cite{30rs} and to a liquid of \Index{linear
molecules} and \Index{arbitrary molecules} by Scheidsteger and
the author \cite{31rs} and by Fabbian et al.~\cite{15rs},
respectively. This was accomplished by the use of the
 \Index{tensorial formalism} which allows the separation of translational
and rotational degrees of freedom. Alternatively, one can also use
a \Index{site-site representation} for molecular systems. MCT
based on such a description was worked out by Chong and Hirata
\cite{32rs} and Chong, G\"otze and Singh \cite{33rs}.

Before discussing MCT let us anticipate that the derivation of the
MCT-equations requires some more or less strong approximations.
Although, these approximations can not be controlled, e.g.~due to
the lack of a smallness parameter, it is interesting to notice
that the MCT-equations for a simple liquid were also obtained by
quite different approaches which are the use of
 \Index{generalized fluctuating hydrodynamics} by Kirkpatrick \cite{34rs}
and Das and Mazenko \cite{35rs}, \Index{density functional
theory} by Kirkpartick and Wolynes \cite{36rs} and recently by the
use of the equation of motion for the microscopic density $\rho
(\vec{q}, t)$ in conjunction with assuming the density
fluctuations to be Gaussian by Zaccarelli et al.~\cite{37rs}. That
these quite different approaches lead to the same mathematical
structure of the MCT-equations shows its ``robustness''. In
addition, it has been argued that the MCT-equations become even
exact in infinite dimensions \cite{36rs}. Further support for MCT
comes from a \Index{spherical model} (without quenched disorder)
\cite{38rs} and for \Index{spin glass models} with infinite range
interactions (see Ref. \cite{8rs,39rs} and their references) for
which the corresponding $q$-independent MCT-equation has been
proven to be exact.

Now we will turn to the discussion of MCT. This will be done in
two sections. In the first one we present the derivation of the
MCT-equations and in the second one the properties of their
solutions.

\subsection{Derivation of the MCT-equations}

We will restrict ourself to a simple liquid of $N$ identical
particles with mass $m$ in a finite volume $V$. We assume two-body
interactions $\upsilon(\vec{x}-\vec{x}')$ such that the
\textit{classical} hamiltonian is given by:

\begin{equation} \label{eq24rs}
H (\{\vec{x}_i\} ,\{\vec{p}_i\})=\sum\limits_n \frac{1}{2 m}
\vec{p}_n^2 + V ( \{\vec{x}_i\})
\end{equation}

\noindent with the potential energy:

\begin{equation} \label{eq25rs}
V(\{\vec{x}_i\} )= \frac{1}{2} \sum\limits_{i \neq j} \upsilon
(\vec{x}_i-\vec{x}_j)
\end{equation}

\noindent $\{\vec{x}_i\}$ and $\{\vec{p}_i \}$ are the
corresponding positions and momenta, respectively. Fixing an
initial point $\{\vec{x}_i\}$, $\{\vec{p}_i\}$ in phase space, the
phase space point $\{\vec{x}_i(t)\},$  $\{\vec{p}_i (t)\}$ at time
$t$ is determined by Newtonian's equation of motion or
equivalently by the hermitean Liouville operator:

\begin{equation} \label{26rs}
\vec{x}_i (t) = e^{i {\cal{L}} t} \vec{x}_i \quad , \quad
\vec{p}_i (t) = e^{i {\cal{L}} t} \vec{p}_i \quad .
\end{equation}

\noindent Of course, one can not solve these equations of motion
for a macroscopic system. Therefore one has to use a theoretical
framework which restricts itself on the relevant variables. For a
liquid this is the 
\index{microscopic density}\textit{microscopic} density:

\begin{equation} \label{eq27rs}
\rho (\vec{x}, t) = \sum\limits_n \delta (\vec{x} - \vec{x}_n (t))
\end{equation}

\noindent where the dependence of the positions $\vec{x}_n(t)$ on
the initial point $\{\vec{x}_i\}$, $\{\vec{p}_i \}$ is suppressed.
Now we recall that $\rho(\vec{x}, t)$ will have a slowly varying
part in the strongly supercooled regime, besides fast motions
(vibrations) around the quasi-equilibrium positions. Therefore, we
can consider $\rho(\vec{x}, t)$ or its Fourier transform

\begin{equation} \label{eq28rs}
\rho (\vec{q}, t) = \sum_n \, e^{i \vec{q} \vec{x}_n (t)} = e ^{i
{\cal L} t} \, \rho (\vec{q})
\end{equation}

\noindent as a 
\index{slow variable}\textit{slow variable}.
If $\rho (\vec{q},
t)$ is slow, then the \Index{current density}\\

$$\vec{j} (\vec{q}, t) = \sum\limits_n \, \frac{1} {m} \vec{p}_n
(t) e^{i \vec{q} \vec{x}_n (t)}$$

\noindent is slow, too, since it is related to $\rho(\vec{q}, t)$
through the continuity equation:

\begin{equation} \label{eq29rs}
\dot{\rho}(\vec{q}, t) \equiv i {\cal L} \rho (\vec{q}, t) = i
\vec{q} \cdot \vec{j} (\vec{q}, t) \quad .
\end{equation}

\noindent We can continue taking time derivatives. Then
$\ddot{\rho}(\vec{q}, t)$ is given by:

\begin{equation} \label{eq30rs}
\ddot{\rho} (\vec{q}, t) = i \vec{q} \cdot \sum\limits_n
\frac{1}{m} \, \dot{\vec{p}}_n (t) \, e^{i \vec{q} \vec{x}_n(t)}
\, + \, \textrm{kinetic part} \quad .
\end{equation}

\noindent Using ${\dot{\vec{p}}}_n/m= - \partial V (\{\vec{x}_i\}
)/\partial \vec{x}_n$ and that Eq.~(\ref{eq25rs}) can be

rewritten as

\begin{eqnarray} \label{eq31rs}
V(\{\vec{x}_i\}) = \frac{1}{2} \, \int d^3 x \, \int d^3 x' \, \,
\upsilon(\vec{x}-\vec{x}') \, \rho (\vec{x}) \rho
(\vec{x}')\nonumber\\
=\frac{1}{2V} \, \sum\limits_{\vec{q}} \, \tilde{\upsilon}
(\vec{q}) \rho^* (\vec{q}) \, \rho(\vec{q})
\end{eqnarray}

\noindent  it is easy to prove that:

\begin{equation} \label{eq32rs}
\ddot{\rho} (\vec{q}, t) = \frac{1}{2V} \sum\limits_{\vec{q}_1,
\vec{q}_2}{\!\!\!}^{\prime} w_0 (\vec{q}, \vec{q}_1, \vec{q}_2) \,
\rho (\vec{q}_1, t) \, \rho (\vec{q}_2, t) + \textrm{kinetic part}
\end{equation}

\noindent with the bare and time-independent \Index{vertex}:

\begin{equation} \label{eq33rs}
w_0(\vec{q}, \vec{q}_1, \vec{q}_2) = \,\,\, \vec{q} \cdot
\left[\vec{q}_1 \tilde{\upsilon} (\vec{q}_1) + \vec{q}_2
\tilde{\upsilon} (\vec{q}_2)\right].
\end{equation}

Here we have used that the Fourier transform $\tilde{\upsilon}
(\vec{q})$ of the pair potential is real and $\sum'$ denotes
summation such that $\vec{q}_1+\vec{q}_2= \vec{q}$. The result
Eq.~(\ref{eq32rs}) reveals that the ``force'' $\ddot{\rho}
(\vec{q},t )$ contains contributions from a pair of modes, which
are coupled by the bare vertex $w_0$. It is rather obvious that
the use of a $r$-body interaction would yield a contribution
$\rho(\vec{q}_1, t) \rho (\vec{q}_2, t) \cdot  \ldots \cdot \rho
(\vec{q}_r, t)$. Consequently, the ``force'' $\ddot{\rho}
(\vec{q}, t)$ must be slow, as well. By continuing this procedure
one obtains a set of slow variables. In the following, however, we
will restrict ourself onto the two variables $\rho (\vec{q}, t)$
and $j(\vec{q}, t)= \vec{q} \cdot \vec{j} (\vec{q}, t) /q
(q=|\vec{q}|)$, the \Index{longitudinal current density}. But we
have to keep in mind that $\ddot{\rho} (\vec{q}, t)$ also is slow.
Having chosen $\rho(\vec{q})$ and $j(\vec{q})$ (at $t=0$) as slow
variables one can apply the \Index{Mori-Zwanzig projection
formalism} \cite{40rs,41rs} to derive an \textit{exact} equation
of motion for the normalized \Index{intermediate scattering
function} $\Phi(\vec{q}, t)= S(\vec{q}, t) / S (\vec{q})$ with

\begin{equation} \label{eq34rs}
S(\vec{q}, t) = \frac{1} {N} \langle \rho (\vec{q}, t)^* \rho
(\vec{q}) \rangle =\frac{1} {N} \langle \rho^* (\vec{q}) e^{-i
{\cal L} t} \, \rho (\vec{q}) \rangle
\end{equation}

\noindent where we used the hermiticity of ${\cal L}$.
$S(\vec{q})$ is the \Index{static structure factor} which depends
on the thermodynamical variables $T, n $, etc. The exact
Mori-Zwanzig equation for $\Phi(\vec{q}, t)$ reads:

\begin{equation} \label{eq35rs}
\ddot{\Phi} (\vec{q}, t) + \Omega_q^2 \, \Phi(\vec{q}, t) +
\int\limits_0^t d t' \, M (\vec{q}, t-t') \dot{\Phi} (\vec{q},
t')=0
\end{equation}

\noindent with the \Index{memory kernel}:

\begin{equation} \label{eq36rs}
M(\vec{q}, t) = \frac{1}{N} \,  \frac{m} {k_BT} \, \frac{1}{q^2}
\langle \ddot{\rho} (\vec{q})^* \, Q \, e^{-i Q {\cal L} Q t} \, Q
\ddot{\rho} (\vec{q}) \rangle
\end{equation}

\noindent and the \Index{microscopic frequencies}:

\begin{equation} \label{eq37rs}
\Omega_q =\left(\frac{k_BT}{m} \,\, \frac{q^2}{S(\vec{q})}
\right)^{1/2} .
\end{equation}

The initial condition is $\Phi(\vec{q}, 0)=1$, and $\dot{\Phi}
(\vec{q}, 0)=0$ for all $\vec{q}$. The result Eq.~(\ref{eq35rs})
makes obvious that the problem to calculate $\Phi(\vec{q}, t)$ has
been shifted to the calculation of $M (\vec{q}, t)$, which seems
to be hopeless, as well. But this is not really true. In contrast
to $\Phi$ it is possible to approximate the memory kermel $M$.
$M(\vec{q}, t)$ is the correlation function of the ``forces'' $Q
\, \ddot{\rho} (\vec{q})$, however, with the 
\index{reduced Liouvillian}{\it reduced Liouvillian} 
${\cal L}'= \, Q \, {\cal L} \, Q$. $Q$ is the
projector (see below) which projects perpendicular to both slow
variables $\rho(\vec{q})$ and $j(\vec{q})$. Since $\ddot{\rho}
(\vec{q})$ contains a coupled pair of modes $\rho (\vec{q}_1)\,
\rho (\vec{q}_2)$ (cf.~Eq.~(\ref{eq32rs})) and because $Q
\rho(\vec{q}_1) \, \rho(\vec{q}_2) \neq 0$, the ``force'' $Q
\ddot{\rho}(\vec{q})$ still contains a slow part. This suggests to
use the following approximation \cite{27rs}:

\begin{equation}\label{eq38rs}
Q \ddot{\rho}(\vec{q}) \rightarrow Q |\ddot{\rho}(\vec{q})\rangle
\approx {\cal P}Q|\ddot{\rho}(\vec{q}) \rangle
\end{equation}
\noindent where ${\cal P}$ is the projector onto pairs of modes:
\begin{equation}\label{eq39rs}
{\cal P}= \sum \limits _ {\vec{q}_1
\vec{q}_2,\vec{q}_1'\vec{q}_2'} g(\vec{q}_1 \vec{q}_2;
\vec{q}_1'\vec{q}_2')|\rho (\vec{q}_1)\rho (\vec{q}_2) \rangle
\langle \rho (\vec{q}_1')^* \rho (\vec{q}'_2)^*|
\end{equation}
and $g$ is determined such that ${\cal P}^2= {\cal P}$. The reader
should note that we have introduced a bra- and ket-notation
$\langle |$ and $| \rangle$, respectively, like in quantum
mechanics. This can be done since the canonical average $\langle
A^*B\rangle $ of two phase space functions $A^*$ and $B$ can be
interpreted as scalar product $\langle A^*|B\rangle$. It is the
existence of this scalar product which allows to introduce
projectors. Substituting Eq.~(\ref{eq38rs}) and Eq.~(\ref{eq39rs})
into Eq.~(\ref{eq36rs}) relates $M(\vec{q}, t)$ to the correlation
function
\begin{equation}\label{eq40rs}
\langle \rho (\vec{q}_3')^* \rho (\vec{q}_4') ^* |e^{-i{\cal L}'t}
|\rho (\vec{q}_1) \rho (\vec{q}_2)\rangle  \quad .
\end{equation}
This relationship involves $g (\vec{q}_1 \vec{q}_2;
\vec{q}_1'\vec{q}_2') g (\vec{q}_3 \vec{q}_4;
\vec{q}_3'\vec{q}_4')$ and $\langle \ddot{\rho}
(\vec{q})^*Q|\rho (\vec{q}_3)\rho (\vec{q}_4)\rangle \\
\langle \rho (\vec{q}_1')^* \rho (\vec{q}_2')^* |Q \ddot{\rho}
(\vec{q}_1)\rangle $ which are \textit{static} quantities. Now,
the crucial approximation is the factorization of the correlator
Eq.~(\ref{eq40rs}) and simultaneously replacing ${\cal L}'$ by
${\cal L}$:

\begin{eqnarray}\label{eq41rs}
\langle \rho (\vec{q}'_3)^* \rho (\vec{q}'_4)^*| e^{- i {\cal L}'
t} | \rho (\vec{q}_1)  \rho (\vec{q}_2) \rangle \approx
\nonumber\\
\approx [\langle \rho (\vec{q}_1)^*| e^{-i {\cal L} t}
| \rho (\vec{q}_1 ) \rangle \, \langle \rho(\vec{q}_2)^*|e^{- i
{\cal L} t} | \rho(\vec{q}_2) \rangle
\delta_{q'_3 q_1} \, \delta_{q'_4 q_2} + \nonumber\\
+ (1 \longleftrightarrow 2)]\nonumber\\
=N^2 S(\vec{q}_1) S (\vec{q}_2) \,\, \Phi(\vec{q}_1, t) \,
\Phi(\vec{q}_2, t) [ \delta_{q'_3 q_1} \, \delta_{q'_4 q_2} +
\delta_{q'_3 q_2} \, \delta_{q'_4 q_1}] \quad .
\end{eqnarray}

The condition ${\cal P}^2 = {\cal P}$ implies that $g(\vec{q}_1
\vec{q}_2; \vec{q}'_1\vec{q}'_2)$ is the inverse of the ``matrix''
$(\langle \rho (\vec{q}'_1)^* \rho (\vec{q}'_2)^* \rho
(\vec{q}_1)\rho (\vec{q}_2)\rangle )$. Since this ``matrix''
equals the correlator Eq.~(\ref{eq40rs}) at $t=0$ it is
approximated by the r.h.s. of Eq.~(\ref{eq41rs}), at $t=0$. Using
this approximation one immediately finds:

\begin{equation}\label{eq42rs}
g(\vec{q}_1, \vec{q}_2; \, \vec{q}'_1, \vec{q}'_2) \approx (4 N^2
S (\vec{q}_1) S (\vec{q}_2))^{-1} [\delta_{q'_1 q_1} \,
\delta_{q'_2 q_2} + \delta_{q'_1 q_2} \, \delta_{q'_2 q_1}] \quad.
\end{equation}

Using $Q=1-{\cal P}=1-(N \, S(\vec{q}))^{-1}| \rho(\vec{q})
\rangle \langle \rho (\vec{q})^*| - (N k_B T/m)^{-1} | j (\vec{q})
\rangle \langle j (\vec{q})^*|$ the static correlation $\langle
\ddot{\rho} (\vec{q})^* \, Q| \rho (\vec{q}_1) \rho (\vec{q}_2)
\rangle$ can be expressed as follows \cite{27rs}:

\begin{eqnarray} \label{eq43rs}
&&\langle \ddot{\rho} (\vec{q})^* Q | \rho (\vec{q}_1) \rho
(\vec{q}_2) \rangle= N \frac{k_B T}{m} \, S(\vec{q}_1) \,
S(\vec{q}_2) \{ - n \vec{q} \cdot [\vec{q}_1 c (\vec{q}_1) +
\vec{q}_2 c (\vec{q}_2) ] + \nonumber\\
&&\qquad \qquad {}+ q^2 [1- \langle \rho (\vec{q})^* \rho
(\vec{q}_1) \rho (\vec{q}_2) \rangle / (N \, S (\vec{q}) \, S
(\vec{q}_1) \, S(\vec{q}_2))]\} \, \delta_{q_1+q_2,q}
\end{eqnarray}

\noindent where we introduced the \Index{direct correlation
function} $c(\vec{q})$ defined by $S(\vec{q})=[1-n c
(\vec{q})]^{-1}$. Substituting Eqs.~(\ref{eq38rs})-(\ref{eq43rs})
into Eq.~(\ref{eq36rs}) yields finally the
 \Index{MCT-approximation} for the memory kernel:

\begin{equation} \label{eq44rs}
M(\vec{q}, t) \approx \nu_q \, \delta (t) + \Omega_q^2 m (\vec{q},
t)
\end{equation}

\begin{equation} \label{eq45rs}
m (\vec{q}, t) = \frac{1} {2 V} \, \sum\limits_{\vec{q}_1,
\vec{q}_2}{\!\!\!}' \, V (\vec{q}, \vec{q}_1, \vec{q}_2) \, \Phi
(\vec{q}_1, t) \, \Phi (\vec{q}_2, t)
\end{equation}

\noindent with the \textit{positive} \Index{vertices}:

\begin{eqnarray} \label{eq46rs}
V (\vec{q}, \vec{q}_1, \vec{q}_2) = \frac{1} {n} \, S (\vec{q}) \,
S (\vec{q}_1) \, S(\vec{q}_2) \, q^{-4} \{ n \vec{q} \cdot
[\vec{q}_1 c (\vec{q_1}) + \vec{q}_2 c (\vec{q}_2)] - \nonumber\\
-q^2 [1- \langle \rho (\vec{q})^* \rho (\vec{q}_1) \rho
(\vec{q}_2) \rangle / (N \, S(\vec{q}) \, S(\vec{q}_1) \,
S(\vec{q}_2))]\}^2 .
\end{eqnarray}

The first term on the r.h.s of Eq.~(\ref{eq44rs}) accounts for the
fast part of the ``force'' $Q \ddot{\rho} (\vec{q})$ leading to a
frictional contribution.

The equations Eqs.~(\ref{eq35rs}),~(\ref{eq37rs}) and
~(\ref{eq44rs})-~(\ref{eq46rs}) are the 
\index{mode coupling equations}{\it mode coupling equations}. 
Due to the MCT-approximation they are a
\textit{closed} set of integro-differential equations for the
normalized correlator $\Phi(\vec{q}, t)$ with initial conditions
$\Phi(\vec{q}, 0) \equiv 1$ and $\dot{\Phi}(\vec{q}, 0)\equiv 0$,
because of time reversal symmetry. As an input they only need the
\textit{static} two-point correlator $S(\vec{q})$ (or equivalently
the direct correlation function $c (\vec{q}$)), the
\textit{static} three point correlator $\langle \rho (\vec{q})^*
\rho (\vec{q}_1) \, \rho(\vec{q}_2) \rangle$ and the frictional
constants $\nu_q$. $\nu_q$ which can only be determined from
kinetic theory does not influence the glassy behavior. Therefore
it can be put to zero. The remaining \textit{static} two- and
three point correlators $S(\vec{q} )= \langle \rho (\vec{q})^*
\rho (\vec{q}) \rangle / N$ and $\langle \rho (\vec{q})^* \, \rho
(\vec{q}_1) \, \rho (\vec{q}_2) \rangle$ can be calculated for
given potential energy $V( \{ \vec{x}_i \})$. This is what makes
MCT a microscopic first principle theory. Here a comment is in
order: MCT will be applied to the supercooled liquid, i.e.~to a
temperature regime where the stable thermodynamical phase is a
crystal. In order to study the glass transition one has to use the
static correlators for the supercooled liquid and \textit{not} for
the crystal. MCT allows to predict the time dependence of the
density fluctuations provided both static correlators are known.
They can be obtained either from analytical approximation schemes
\cite{40rs} or from experiments and simulations. Application of
the \Index{convolution approximation} \cite{40rs}:

\begin{equation} \label{eq47rs}
\langle \rho (\vec{q})^* \rho (\vec{q}_1) \rho (\vec{q}_2) \rangle
\approx N \, \, S(\vec{q}) S(\vec{q}_1) S(\vec{q}_2) \, \,
\delta_{q_1 + q_2, q}
\end{equation}

\noindent leads to a further simplification of the vertices
Eq.~(\ref{eq46rs}):

\begin{equation} \label{eq48rs}
V (\vec{q}, \vec{q}_1, \vec{q}_2) = n  S (\vec{q}) S(\vec{q_1})
S(\vec{q}_2) \, [\vec{q} \cdot (\vec{q}_1 c (\vec{q}_1) +
\vec{q}_2 c(\vec{q_2}))]^2 / q^4
\end{equation}

\noindent which involve $S(\vec{q})$ (or $c(\vec{q})$), only. For
SiO$_2$-liquids it has been demonstrated by Sciortino and Kob
\cite{42rs} that a satisfactory agreement of, e.g.~the critical
nonergodicity parameters determined from a MD-simulation with
those from MCT is only obtained with the vertices from
Eq.~(\ref{eq43rs}) where the three-point correlator is
\textit{not} factorized. This is rather plausible, because SiO$_2$
is a covalent glass former with bond orientational correlations
which are completely neglected by the convolution approximation
Eq.~(\ref{eq47rs}).

\subsection{Solutions and predictions of MCT}

The main question which arises is: How can one detect a glass
transition within MCT? The answer is rather simple. Let us use the
\index{nonergodicity parameter}{\it nonergodicity parameters} 
defined by:
\begin{equation}\label{eq49rs}
f(\vec{q})= \lim \limits _{t \rightarrow \infty} \Phi (\vec{q},t)=
- \lim \limits _{z \rightarrow \infty} z \hat{\Phi} (\vec{q},z)
\quad .
\end{equation}
These parameters, which are just the infinite time limit of $\Phi
(\vec{q},t)$ or the zero frequency limit of its Laplace transform
\begin{equation}\label{eq50rs}
\hat{\Phi}(\vec{q},z)= i \int \limits _0 ^\infty dt \Phi
(\vec{q},t)e^{izt} \quad , \; \textrm{Im} \, z>0
\end{equation}
can be used as \Index{glass order parameters}, because they
vanish in an \Index{ergodic} phase (= liquid phase) and are
nonzero in a \Index{nonergodic} one, which is interpreted as a
glass phase. To be more precise, this is true for the correlation
function of the density fluctuations $\delta \rho (\vec{q}, t) =
\rho (\vec{q}, t)- \langle \rho (\vec{q}, t) \rangle$, only. Since
$\delta \rho (0,t)=0$ and $\delta \rho (\vec{q}, t)= \rho
(\vec{q}, t)$ for all $\vec{q} \neq 0$, it is sufficient to
investigate $\Phi(\vec{q}, t)$. Thus it is:
\begin{equation}\label{eq51rs}
f(\vec{q}) = \left \{ \begin{array}{r@{\quad , \quad}l} 0 &
\textrm{liquid} \\ > 0 & \textrm {glass}
\end{array} \right.
\end{equation}
Since $\ddot{\Phi}(\vec{q},t)\rightarrow 0$ for $t \rightarrow
\infty$, it is easy to show that Eq.~(\ref{eq35rs}) yields a
nonlinear set of algebraic equations for $f(\vec{q})$:
\begin{equation}\label{eq52rs}
\frac{f(\vec{q})}{1-f (\vec{q})}={\cal F}[f(\vec{q})]
\end{equation}
with:
\begin{equation} \label{eq53rs}
{\cal F} [f(\vec{q})]= \frac {1}{2V} \sum
\limits_{\vec{q}_1,\vec{q}_2}{\!}^{\prime} \, V(\vec{q},
\vec{q}_{1},\vec{q}_2) f(\vec{q}_1)f(\vec{q}_2) \quad .
\end{equation}
One can prove that the long time limit $f(\vec{q})$ is
distinguished from other possible solutions of
Eq.~(\ref{eq52rs}),~(\ref{eq53rs}) by the following properties
\cite{27rs}

\begin{itemize}
\item[{(i)}]
\hspace{0,02cm}$ f(\vec{q})$ is real (since $\Phi (\vec{q},t)$ is
real)

\item[{(ii)}]
\hspace{0,02cm}$0 \leq f(\vec{q}) \leq 1$

\item[{(iii)}]
\hspace{0,03cm}If several solutions $f_1(\vec{q}) > f_2(\vec{q})
\geq f_3(\vec{q}) \geq \cdots $ exist, the long time limit is the
largest one: $f(\vec{q})=f_1(\vec{q})$ .
\end{itemize}

The positiveness of the vertices is crucial for (iii).

It is obvious from Eqs.~(\ref{eq52rs}),~(\ref{eq53rs}) that
$f(\vec{q}) \equiv 0$ is always a solution. In order to check
whether a nontrivial solution exists, let us neglect the
$q$-dependence for a moment. This leads to a so-called
 \Index{schematic model} \cite{27rs}, the ${\cal F}_2$-model, where ${\cal
F} [f]= \upsilon f^2, \upsilon \geq 0$, i.e.~we have to solve
\begin{equation}\label{eq54rs}
\frac{1}{1-f} = vf^2 \quad .
\end{equation}
For $f \neq 0$ this leads to a quadratic equation with solution:
\begin{equation} \label{eq55rs}
f_{1/2}= \frac 1 2 [1 \pm \sqrt{1-4/v} ] \quad .
\end{equation}

\begin{figure}[t]
\unitlength1cm
\begin{picture}(11,6)
\centerline{\psfig{file=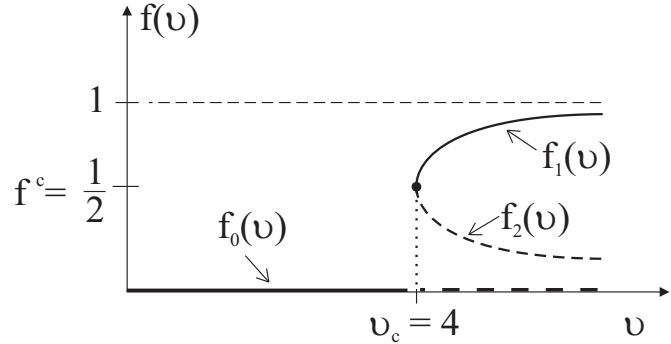,angle=0,width=12cm}}
\end{picture}
\caption{ \label{fig3rs} Qualitative control parameter dependence
of the nonergodicity parameter of the ${\mathcal{F}}_2$-model.
$f_0(v) \equiv 0$ is the trivial solution. At $v_c=4$ two new
solutions $f_1(v)$ and $f_2(v)$ bifurcate. }
\end{figure}

Since $f_{1/2}$ is not real for $v <v_c=4$, the only physical
solution in this range is $f_0 =0$. For $v \geq v_c$ two real
solutions $f_1 > f_2$ bifurcate from the trivial one. This
scenario is illustrated in Fig.~\ref{fig3rs}. Hence, we get for
the $F_2$-model
\begin{equation}\label{eq56rs}
f(v)=\left \{ \begin{array}{r@{\quad , \quad}l} 0 & v <v_c \\
\frac 1 2 [1+\sqrt{1-v_c/v}] & v \geq v_c
\end{array}\ \right.
\end{equation}
i.e. at the critical control parameter $v_c=4$ the nonergodicity
parameter changes \textit {discontinuously} from zero to the
critical nonergodicity parameter $f^c=f(\upsilon_c) = 1/2$. This
type of transition is called 
\index{type-B transition}{\it type-B transition}, in
contrast to a \Index{type-A transition} which is {\it continuous}
\cite{27rs}. Such a {\it type-A transition} occurs, e.g. for the
${\cal F}_1$-model: ${\cal F}[f]= \upsilon f$, $\upsilon \geq 0$.

This simple example has shown that an ergodic-to-nonergodic
transition can occur at a critical coupling constant $v_c$. Since
the vertices $V (\vec{q},\vec{q}_1,\vec{q}_2)$ are positive and $0
\leq f(\vec{q}) \leq 1$, ${\cal F} [f (\vec{q})]$ will diverge in
the strong coupling limit $V
(\vec{q},\vec{q}_1,\vec{q}_2)\rightarrow \infty $. Therefore
Eq.~(\ref{eq52rs}) can only be fulfilled if the nontrivial
solution converges to one. Therefore there must be a critical
hypersurface in the control parameter space at which transitions
from $f(\vec{q})\equiv 0$ to $f(\vec{q})\neq 0$ happen. Because
the vertices depend through $S(\vec{q})$ on the thermodynamic
variables $T,n$ etc. and increase with decreasing $T$ (increasing
$n$) there will be a \Index{critical temperature} $T_c$ (critical
density $n_c$) at which the system undergoes an
ergodic-to-nonergodic transition, i.e. a glass transition. From
this we can conclude that MCT yields a {\it dynamical} glass
transition whereas static quantities, e.g. $S(\vec{q})$, are
\textit{not} singular at $T_c$ (or $n_c$). Having established the
existence of a glass transition singularity, one can study the
dynamics close to it. For large times (small frequencies) compared
to the microscopic time scale $\Omega_q^{-1} (\Omega_q)$ the
Laplace transform of Eq.~(\ref{eq35rs}) yields with
Eq.~(\ref{eq44rs}):

\begin{equation} \label{eq57rs}
\frac{z \hat{\Phi} (\vec{q}, z)} {1 + z \hat{\Phi} (\vec{q}, z)} =
z \hat{m} (\vec{q}, z) \quad .
\end{equation}

\begin{figure}[t]
\unitlength1cm
\begin{picture}(11,5.5)
\centerline{\psfig{file=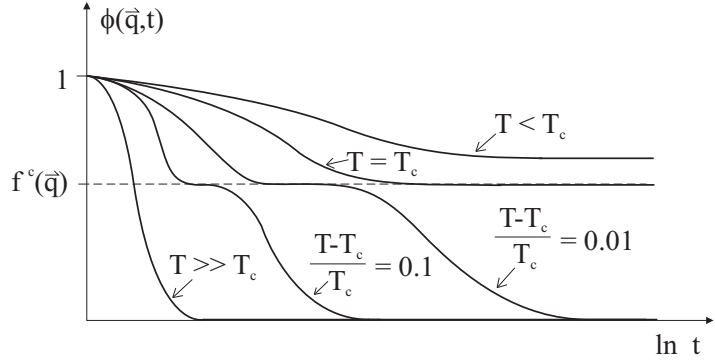,angle=-90,width=13cm}}
\end{picture}
\caption{ \label{fig4rs} Qualitative time- and
temperature-dependence of the intermediate scattering function
$\Phi(\vec{q},t)$ for fixed $\vec{q}$. The values 0.1 and 0.01 for
($T-T_c)/T_c$ should not be taken literally. They should only
indicate that $T$ is chosen closer and closer to $T_c$.}
\end{figure}

Here we used $\nu_q=0$. The reader should notice that
Eq.~(\ref{eq57rs}) does not involve anymore the microscopic
frequencies $\Omega_q$. The schematic representation of a solution
of the MCT-equations
Eqs.~(\ref{eq35rs}),~(\ref{eq44rs})-~(\ref{eq46rs}), including the
microscopic time regime, is shown in Figure~\ref{fig4rs}. This
Figure clearly demonstrates the existence of a critical
temperature $T_c$ at which a jump of $f(\vec{q})$ occurs. The
generic behavior close to $T_c$ is given by (cf.~also
Eq.~(\ref{eq56rs})):

\begin{eqnarray} \label{eq58rs}
f(\vec{q})= \left \{\begin{array}{ll} 0  \,\,\,\ ,  \,\,\,\  T > T_c\\
f^c (\vec{q}) + \textrm{const} (T_c-T)^{1/2} \, \, , \, \, T \leq T_c\\
\end{array} \right.
\end{eqnarray}

\noindent where the constant is $q$-dependent. We can observe from
Figure~\ref{fig4rs} that \textit{close} to $T_c$ there is a time
scale on which the correlator $\Phi(\vec{q}, t)$ is close to the
critical nonergodicity parameter $f^c (\vec{q})=
f(\vec{q})|_{_{T=T_c}}$. This suggests the Ansatz:

\begin{equation} \label{eq59rs}
\Phi (\vec{q}, t) = f^c (\vec{q}) + h (\vec{q}) G(t)
\end{equation}

\noindent with $|G(t)| \ll 1$. It is important to realize that the
$q$- and $t$-dependence is factorized. This is related to the type
of \Index{bifurcation scenario} for which the {\it
non-degenerated} largest eigenvalue of the stability matrix of the
linearized Eq.~(\ref{eq52rs}) becomes one. Therefore at $T_c$ only
\textit{ one} unstable eigenvector occurs. The \Index{critical
amplitude} $h(\vec{q})$ is the amplitude of that eigenvector.
Substituting Eq.~(\ref{eq59rs}) into Eq.~(\ref{eq57rs}) and
expanding up to quadratic order in $G$ one obtains:

\begin{equation} \label{eq60rs}
\sigma + \lambda \{- z L T [G^2 (t)] (z)\}- \{ -z \widehat{G} (z)
\}^2 =0
\end{equation}

\noindent with the \Index{separation parameter}

\begin{eqnarray} \label{eq61rs}
\sigma(T) \cong \sigma_0 (T_c-T)=
\left \{\begin{array}{ll} < 0  \,\,\,\ ,  \,\,\,\  \textrm{liquid}\\
> 0 \,\,\,\ ,  \,\,\,\ \textrm{glass}\\
\end{array} \right.
\end{eqnarray}

\noindent ($\sigma_0$ is positive) and the \Index{exponent
parameter}

\begin{equation} \label{eq62rs}
\lambda = \lambda [S(\vec{q})|_{_{T=T_c}}]
\end{equation}

The separation parameter is a measure of the distance from the
transition point. The importance of $\lambda$ will become clear
below. Explicit expression for $\sigma$ and $\lambda$ are given in
Ref.~\cite{27rs}.

At the transition point, i.e.~for $\sigma=0$, the solution of
Eq.~(\ref{eq60rs}) is the 
\index{critical law}\textit{critical law}:

\begin{equation} \label{eq63rs}
G(t) \sim t^{-a} \quad .
\end{equation}

\noindent where the exponent $a$ is a solution of:

\begin{equation} \label{eq64rs}
\lambda = \frac{\Gamma^2 (1-a)} {\Gamma (1-2a)}
\end{equation}

\noindent with $\Gamma (x)$ the Gamma function. Since $\lambda$ is
uniquely determined by $S(\vec{q})$ at $T_c$ it can be calculated
microscopically, from which $a$ follows. $a$ is restricted to $0<
a \leq 1/2$. For $a=0$ a \Index{higher order bifurcation}
scenario occurs \cite{27rs}. To solve Eq.~(\ref{eq60rs}) for
$T\neq T_c$ it is convenient to introduce a \Index{correlation
scale} $\sqrt{|\sigma|}$ by:

\begin{equation} \label{eq65rs}
G(t)= \sqrt{|\sigma|} \,\, g_{\pm} (t / t_\sigma) \quad , \quad
\sigma \gtrless  0
\end{equation}

\noindent where the time scale $t_\sigma$ and the master functions
$g_{\pm}$ can be obtained as follows. Introducing
Eq.~(\ref{eq65rs}) into Eq.~(\ref{eq60rs}) and using
$\hat{t}=t/t_\sigma$ and $\hat{z}=t_\sigma z$ leads to:

\begin{equation} \label{eq66rs}
\pm 1/\hat{z} + \lambda L T [g_\pm^2 (\hat{t})] (\hat{z}) +
\hat{z} (\hat{g}_\pm (\hat{z}))^2=0
\end{equation}

\noindent which has the solution:

\begin{equation} \label{eq67rs}
g_\pm (\hat{t}) \sim \hat{t}^{-a}
\end{equation}

\noindent for $\hat{t} \ll 1$, i.e.~ $t \ll t_\sigma$, since $\pm
1/\hat{z}$ can be neglected due to $\hat{z} \gg 1$. Substitution
of Eq.~(\ref{eq67rs}) into Eq.~(\ref{eq65rs}) and taking into
account that $G(t)$ must reduce for $\sigma \rightarrow 0$ to the
\Index{critical correlator} Eq.~(\ref{eq63rs}) allows to fix
$t_\sigma$:

\begin{equation} \label{eq68rs}
t_\sigma (T) \sim |\sigma|^{-\frac{1} {2a}} \sim
|T-T_c|^{-\frac{1}{2a}} \quad , \quad T \lessgtr  T_c \quad .
\end{equation}

For $\hat{t} \gg 1$ one obtains \cite{27rs}:

\begin{eqnarray} \label{eq69rs}
g_\pm (\hat{t}) \cong \left \{\begin{array}{ll}
\sqrt{1-\lambda}  \,\,\,\ ,  \,\,\,\  (+)\\
- B \hat{t}^b \,\,\,\,\, \,\,\,\, \,  ,  \,\,\,\ (-)\\
\end{array} \right.
\end{eqnarray}

\noindent with $B>0$.

Again, the positive exponent $b$ is determined by the exponent
parameter $\lambda$:

\begin{equation} \label{eq70rs}
\lambda= \frac{\Gamma^2 (1+b)}{\Gamma (1+2b)} \quad .
\end{equation}

In the liquid phase (minus sign in Eq.~(\ref{eq69rs})) we obtain,
besides the 
\index{critical law}{\it critical law} 
Eq.~(\ref{eq63rs}) a
second power law, the so-called 
\index{von Schweidler law}{\it von Schweidler law}.
Substituting $g_{-}(\hat{t})$ from Eq.~(\ref{eq69rs}) into
Eq.~(\ref{eq65rs}) leads to

\begin{equation} \label{eq71rs}
G(t) \sim (t/\tau (T))^b
\end{equation}

\noindent with the second time scale:

\begin{equation} \label{eq72rs}
\tau (T) \sim |\sigma|^{-{\frac{1}{2b}}}\,  t_\sigma \sim
(T-T_c)^{-\gamma} \, , \,\,\,\,  T \geq T_c
\end{equation}

\noindent and the exponent

\begin{equation} \label{eq73rs}
\gamma = \frac{1}{2a} + \frac{1} {2b} \quad .
\end{equation}
\begin{figure}[t]
\unitlength1cm
\begin{picture}(11,5.5)
\centerline{\psfig{file=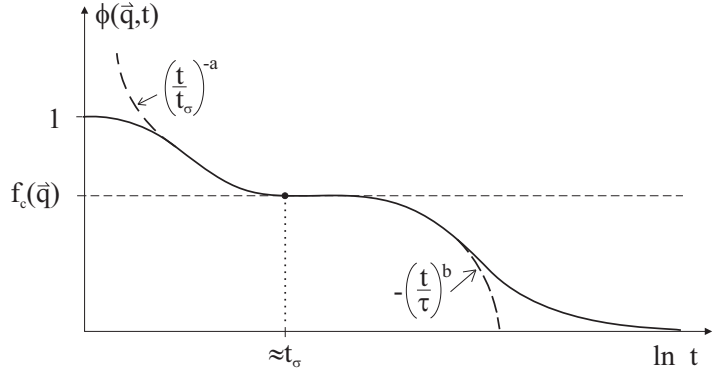,angle=-90,width=13cm}}
\end{picture}
\caption{ \label{fig5rs} $\Phi(\vec{q},t)$ for fixed $\vec{q}$ and
a single temperature $T>T_c$ (solid line). Both power laws, the
critical law and the von Schweidler law, are shown as dashed
lines. The horizontal dashed line is the plateau height equal to
the critical nonergodicity parameter $f_c(\vec{q})$ }
\end{figure}
The von Schweidler law, Eq.~(\ref{eq71rs}), is valid for $\hat{t}
\gg1$, i.e.~$t\gg t_\sigma (T)$ and $t\ll\tau (T)$ because of
$|G(t)| \ll 1$. Both power laws are shown in Figure~\ref{fig5rs}.
The critical law describes the relaxation to the plateau value
$f^c(\vec{q})$ and $f(\vec{q})$ above {\it and} below $T_c$,
respectively, and the von Schweidler law the initial decay from
the plateau for $T > T_c$. Both time scales $t_\sigma$ and $\tau$
exhibit power law divergence at $T_c$. $\tau$ diverges faster than
$t_\sigma$, due to $\gamma > / 1(2a)$ (see Fig.~\ref{fig6rs}).

\begin{figure}[t]
\unitlength1cm
\begin{picture}(11,7)
\centerline{\psfig{file=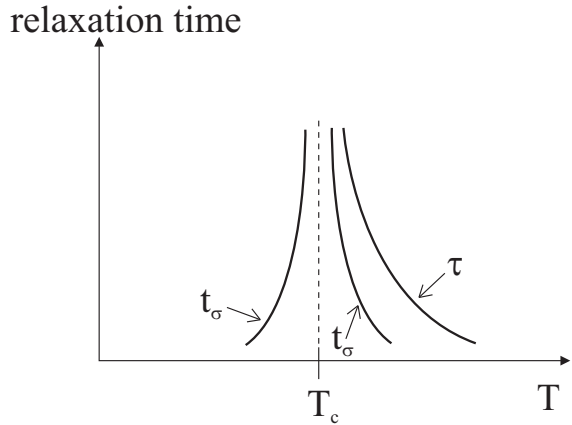,angle=0,width=12cm}}
\end{picture}
\caption{ \label{fig6rs} Qualitative T-dependence of both
relaxation times $t_\sigma (T)$ and $\tau(T)$ which exhibit power
law divergence at $T_c$ }
\end{figure}

These fascinating results proven by G\"otze in 1984 \cite{43rs}
and 1985 \cite{44rs} were absolutely new. The time range $t_0 \ll
t\ll\tau(T)$ ($t_0$ is a microscopic time scale $\sim
\Omega_q^{-1}$) which includes both power laws have been called
(fast) $\beta$-relaxation. It has a simple physical explanation.
If the temperature is low enough each particle feels a cage on a
time scale $t_\sigma$. The first relaxation step (critical law) is
a relaxation within the cage and the second one (von Schweidler
law) is related to the ``opening'' of the cage for $t \gg
t_\sigma$ which is the initial stage of the \Index{structural
relaxation}. The fact that the $q$- and $t$-dependence of the
correlators factorize in the $\beta$-regime is one of the highly
nontrivial predictions of MCT which has not been found before in
condensed matter physics. Numerous experiments \cite{12rs} (see
also the contribution by U. Buchenau in this monograph), and
computer simulations \cite{13rs,14rs,15rs} have given consistent
results with regard to these MCT-predictions.

What remains is to study the dynamics for $t$ of the order or much
larger than $\tau (T)$ which is called 
\Index{$\alpha$-relaxation}. There the factorization does not hold anymore.
Therefore one has to solve Eq.~(\ref{eq57rs}). This can only be
done numerically. But there is one important feature, which is the
\Index{scale invariance} of Eq.~(\ref{eq57rs}). This means that a
scale transformation $t \rightarrow t \cdot y$, $\Phi(\vec{q}, t)
\rightarrow \Phi(\vec{q}, t \cdot y)$ or $z \rightarrow z \cdot
y^{-1}$ , $\hat{\Phi} (\vec{q}, z) \rightarrow y \hat{\Phi}
(\vec{q}, z \cdot y^{-1})$ leaves Eq.~(\ref{eq57rs}) invariant.
This allows to introduce a $q$-dependent master function $\Phi_q
(\tilde{t})$ such that

\begin{equation} \label{eq74rs}
\Phi(\vec{q}, t; T) = \Phi_q (t/\tau(T))
\end{equation}

i.e.~the time and temperature dependence of $\Phi$ appears only in
the combination $t/\tau(T)$. Eq.~(\ref{eq74rs}) is well-known in
the glass science as the \Index{time-temperature superposition
principle}. The master function $\Phi_q(\tilde{t})$ can be
determined from a numerical solution of Eq.~(\ref{eq57rs}). The
validity of Eq.~(\ref{eq57rs}) and Eq.~(\ref{eq74rs}) has also
been test experimentally \cite{12rs} and by computer simulations
\cite{13rs,15rs}.

\section{MICROSCOPIC THEORY: REPLICA THEORY}
In 1996 Mézard and Parisi \cite{45rs} have presented a
 \Index{replica theory} for the structural glass transition. This theory
has been inspired by spinglass theory \cite{1rs,2rs}. Kirkpatrick,
Thirumalai and Wolynes first noticed the analogy between spinglass
and structural glass transitions \cite{8rs}. In 1987 they have
shown that mean-field spin glass models with a \Index{one-step
replica symmetry breaking} \cite{2rs} exhibit a \textit{dynamical}
(MCT-like)glass transition at $T_c$ and a \textit {static} one at
$T_s$, which is below $T_c$. At $T_s$ the \Index{configurational
entropy} vanishes. Hence $T_s$ can be identified with $T_K$, the
\Index{Kauzmann temperature}. In the next section we will
describe the physical picture behind this theory and its
thermodynamic formulation. The second section then contains the
microscopic theory which can be considered as a first principle
approach of an earlier attempt by Singh, Stoessel and Wolynes to
describe the glass transition by a \Index{density functional
theory} \cite{46rs}.

\subsection{Thermodynamical description}
\begin{figure}[t]
\unitlength1cm
\begin{picture}(11,8)
\centerline{\psfig{file=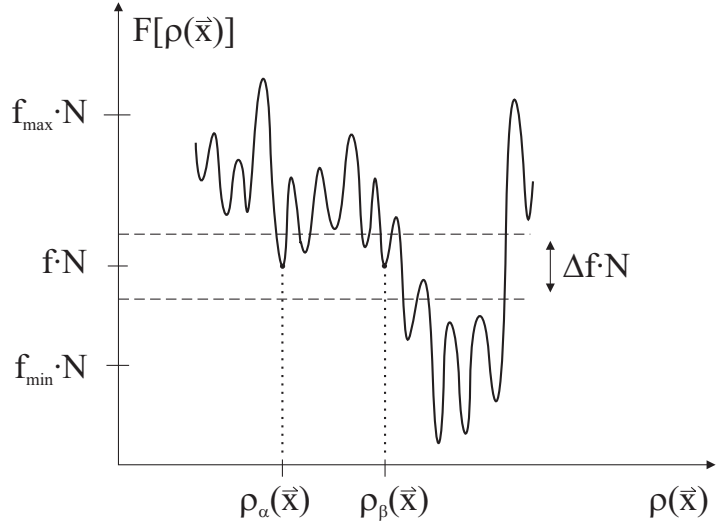,angle=0,width=11cm}}
\end{picture}
\caption{ \label{fig7rs} Schematic illustration of the free energy
landscape. For $f_{\textrm{min}}<f<f_{\textrm{max}}(T)$ there are
exponentially many local minima in an energy interval of width
$\Delta f$ (two of them, $\rho _\alpha(\vec{x})$ and $\rho_\beta
(\vec{x})$, are shown explicitly). Above $f_{\textrm{max}}(T)$ and
below $f_{\textrm{min}}(T)$ there are at most an algebraic number
of minima. }
\end{figure}

Let $\rho (\vec{x})$ be the local particle density of a liquid
with $N$ identical particles in a volume $V$. A theorem
\cite{47rs} guarantees that there exists a \Index{free energy
functional} ${\mathcal{F}}[\rho(\vec{x}),T]$ such that the
equilibrium phases are obtained from
\begin{equation} \label{eq75rs}
\frac {\delta {\mathcal{F}[\rho(\vec{x}),T]}}{\delta
\rho(\vec{x})} =0
\end{equation}
${\mathcal{F}}[\rho(\vec{x}),T]$ is not known exactly. In many
cases one uses an approximation suggested by Ramakrishnan and
Youssouff \cite{48rs}. Let $\rho^{(\alpha)}(\vec{x})$
(T-dependence is suppressed), $\alpha = 1,2,3,\ldots $ be
solutions of Eq.~(\ref{eq75rs}) (local minima!) with free energy
per particle $f_\alpha (T) = {\mathcal{F}} [\rho^{(\alpha)},T]/N$
and let ${\mathcal{N}}(f,N,T)$ be the number of solutions with
free energy $f={\cal F}/N$ at $T$. The configurational entropy per
particle $S_c(f,T)$ (cf.~1st chapter) is defined by:
\begin{equation}\label{eq76rs}
{\mathcal{N}}(f,N,T)= \exp[N S_c(f,T)] \quad .
\end{equation}
At high temperature the equilibrium phase is given by the uniform
density solution $\rho(\vec{x}) \equiv n = N/V$ of
Eq.~(\ref{eq75rs}). Inspired by mean field spin glasses with a
discontinuous transition, Mézard and Parisi \cite{49rs,50rs}
assume that at the MCT-temperature $T_c$ an exponential number of
solutions $\rho^{(\alpha)}(\vec{x})$ occur for $f$ between
$f_{\textrm{min}}(T)$ and $f_{\textrm{max}}(T)$, i.e.
\begin{equation}\label{eq77rs}
S_c(f,T)>0 \; , \quad f_{\textrm{min}}(T) < f < f_{\textrm{max}}
(T) \quad .
\end{equation}
Above $f_{\textrm{max}}(T)$ and below $f_{\textrm{min}}(T)$ it is
$S_c(f,T)=0$. This situation is illustrated in Figure
\ref{fig7rs}. $S_c(f,T)$ varies smoothly with $T$ and is concave
in $f$, i.e.~ $\partial ^2S_c(f,T)/\partial f^2<0$. The
\textit{crucial} assumption is that $S_c(f;T)$ vanishes at
$f_{\textrm{min}}(T)$ with \textit{finite} slope:
\begin{equation}\label{eq78rs}
\frac{\partial S_c}{\partial f}(f_{\textrm{min}},T)< \infty
\end{equation}
(see Figure \ref{fig8rs}). At low temperatures (below $T_c$) the
partition function $Z(T,N)$ can be written as a sum over the
individual local minima:
\begin{equation}\label{eq79rs}
Z(T,N)= \exp [- \beta N \Phi (T)] \cong \sum \limits _\alpha
\exp[- \beta Nf_\alpha (T,N)]
\end{equation}
which becomes for $N$ large
\begin{equation}\label{eq80rs}
Z(T,N) \approx \int \limits ^{f_{max}(T)} _{f_{min}(T)} df \exp
\{-N[\beta f - S_c(f,T)]\}b \quad .
\end{equation}
The main contribution to this integral comes from the minimum
solution $f^*(T)$ of the \Index{free energy}
\begin{equation}\label{eq81rs}
\Phi (f,T)=f-TS_c(f,T) \quad ,
\end{equation}
i.e.
\begin{equation}\label{eq82rs}
\Phi (T) = \min \limits _f \Phi (f,T)= f^*(T)-TS_c(f^*(T),T)\quad
.
\end{equation}
The Boltzmann constant is put to one. Now, there are two
possibilities. \textit{First}, for temperatures below $T_c$ but
high enough $f^*(T)$ will be within the interval
$[f_{\textrm{min}}(T), f_{\textrm{max}}(T)]$. Then $f^*(T)$
follows from the solution of $\partial \Phi (f,T)/\partial f =0$.
Using Eq.~(\ref{eq81rs}) this yields:
\begin{equation}\label{eq83rs}
\frac 1 T = \frac{\partial S_c}{\partial f} (f,T)
\end{equation}
(see Fig.~\ref{fig8rs}a). \textit {Second}, $f^*(T)$ will decrease
with decreasing $T$ and will get stuck at $f_{\textrm{min}}(T)$
(see Fig.~\ref{fig8rs}b). Then it is $S_c(f^*(T),T)=0$ and
\begin{equation}\label{eq84rs}
\Phi(T)=f_{\textrm{min}}(T).
\end{equation}
If we denote by $s_0(T)$ the slope of $\partial S_c/\partial f$ at
$f_{\textrm{min}}(T)$ :
\begin{figure}[t]
\unitlength1cm
\begin{picture}(11,6.5)
\centerline{\psfig{file=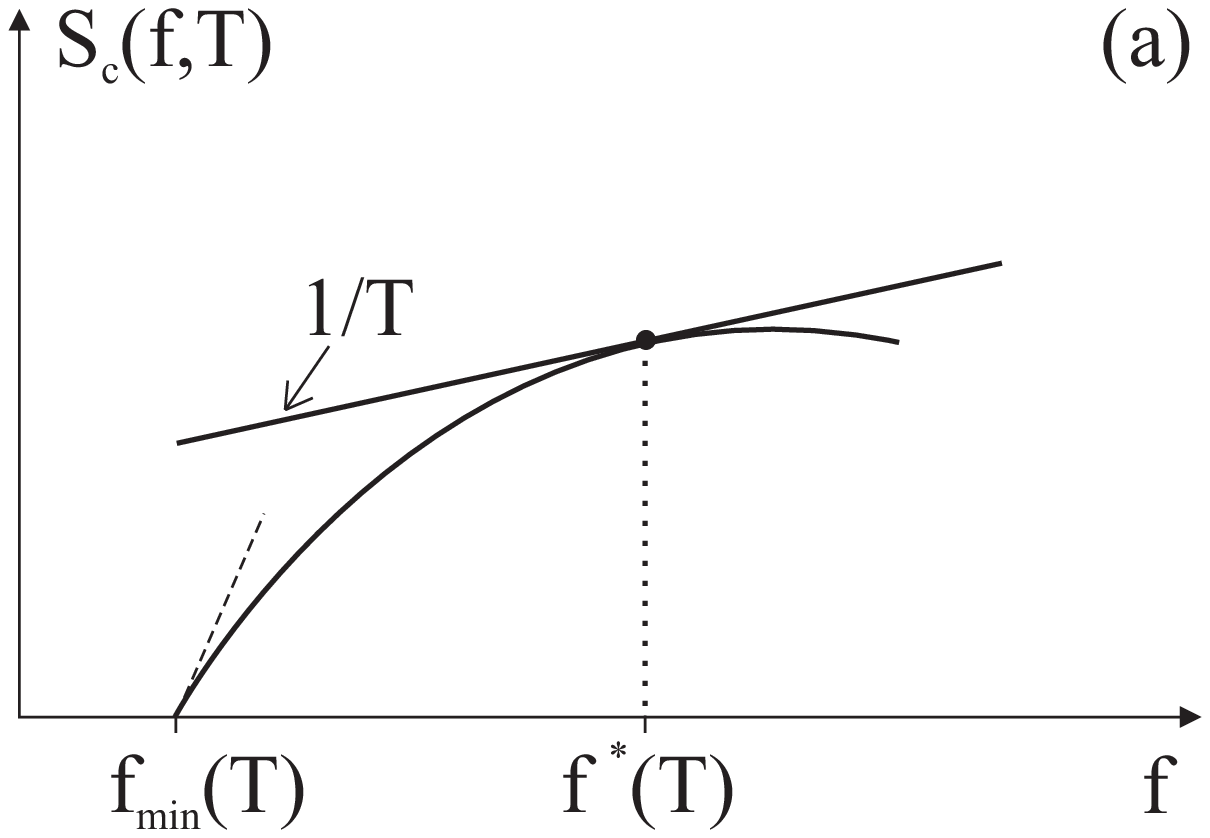,angle=0,width=11cm}}
\end{picture}
\begin{picture}(11,6.5)
\centerline{\psfig{file=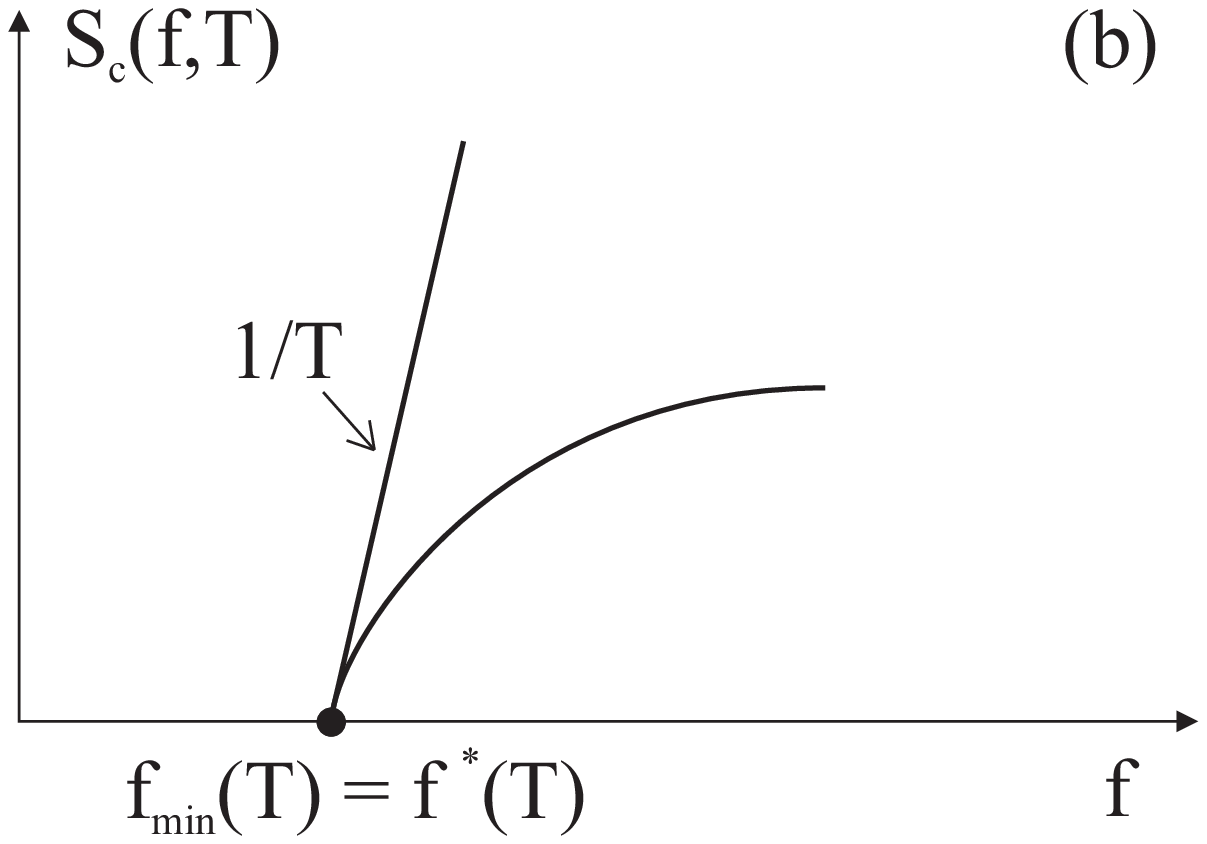,angle=0,width=11cm}}
\end{picture}
\caption{ \label{fig8rs} Schematic $f$-dependence of the
configurational entropy $S_c(f,T)$. (a) $T_K<T<T_c$ where $f^*(T)$
is between $f_{\textrm{min}}(T)$ and $f_{\textrm{max}}(T)$, (b) $T
\leq T_K$ where $f^*(T)$ coincides with $f_{\textrm{min}}(T)$.
Note that the slope of the tangent (which must equal $1/T$) in (a)
is smaller than in (b). The dashed line in part (a) shows the
slope of the tangent at $f_{\textrm{min}} (T)$.}
\end{figure}

\begin{equation} \label{eq85rs}
s_0(T)=\frac{\partial S_c}{\partial f} (f_{\textrm{min}}(T),T) <
\infty
\end{equation}
this will happen at a temperature $T_s > 0$ with
\begin{equation}\label{eq86rs}
\frac {1}{T_s} = s_0(T_s) \quad .
\end{equation}
Since $S_c$ vanishes at $T_s$ it is identical to the
 \Index{Kauzmann temperature} $T_K$. The reader should note that the
existence of a nonzero Kauzmann temperature and accordingly the
existence of a static glass transition at $T_K$ relies on the
finite value of $s_0(T)$ for \textit {all} temperatures, including
zero.

\begin{figure}[t]
\unitlength1cm
\begin{picture}(11,7)
\centerline{\psfig{file=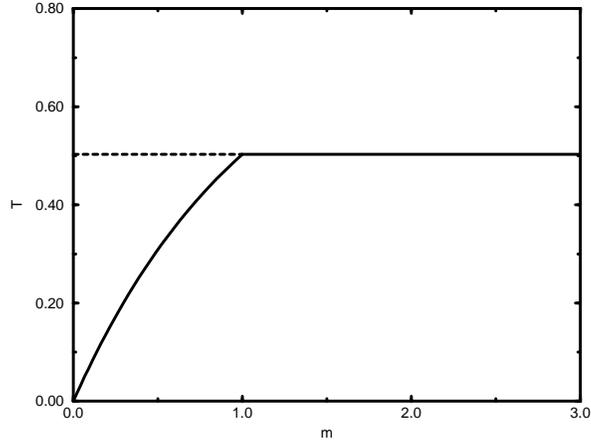,angle=-90,width=9cm}}
\end{picture}
\caption{ \label{fig9rs} Phase diagram of the replica system in
the m-T-space. Solid line separates the liquid from the glass
phase (see also text). The dashed line within the liquid phase
separates the ``atomic'' liquid for $T>T_K \simeq 0.5$ from the
``molecular'' liquid for $T<T_K$. The curved part of the solid
line is $m^*(T)$. }
\end{figure}

Now, it is the major goal to calculate the free energy $\Phi (T)$
and the configurational entropy \textit {below} $T_K$. This can be
done by a trick. Instead of taking one system one chooses $m$
\Index{replicas} which are weakly coupled to each other with
coupling constant $\varepsilon$ \cite{49rs,50rs,51rs}. In the
glass phase, i.e. below $T_K$ this small coupling will force all
systems into the \textit {same} local minimum $\alpha$. Therefore
the corresponding partition function is given by:
\begin{equation}\label{eq87rs}
Z_m(T,N) \approx \int \limits ^{f_{max}(T)} _{f_{min}(T)} df \exp
\{-N[m\beta f - S_c(f,T)]\}
\end{equation}
in analogy to Eq.~(\ref{eq80rs}). The ``saddle point'' condition
Eq.~(\ref{eq83rs}) is:
\begin{equation} \label{eq88rs}
\frac m T = \frac {\partial S_c}{\partial f} (f,T) \quad .
\end{equation}
If we allow $m$ to take any real positive value, it is obvious
that Eq.(~\ref{eq88rs}) has a solution $f^*_m(T) >
f_{\textrm{min}}(T)$ if $m <1$ is small enough, even if $f^*_{m=1}
(T)=f_{\textrm{min}}(T)$. Then the free energy $\Phi(m,T)$ is
given by:
\begin{equation}\label{eq89rs}
\Phi(m,T)=\min \limits _f [mf-TS_c(f,T)] \quad .
\end{equation}
Next, it is easy to prove that $\Phi(m,T)$ and the free energy per
particle of the replicated system:
\begin{equation}\label{eq90rs}
\phi(m,T)= \frac 1 m \Phi(m,T)
\end{equation}
allows to calculate $f(T)$ and $S_c(T)$:
\begin{equation}\label{eq91rs}
f(T)= \frac{\partial \Phi (m,T)}{\partial m}
\end{equation}

\begin{equation}\label{eq92rs}
S_c(T)= \frac {m^2}{T}\frac {\partial \phi (m,T)} {\partial m}
\quad .
\end{equation}
Hence, the knowledge of $\Phi(m,T)$ allows to determine $f$ and
$S_c$. Increasing $m$ towards one there will be a critical value
$0 \leq m^*(T) \leq 1$ at which $f_m^*(T) = f_{\textrm{min}}(T)$.
A schematic representation of the $m-T$ phase diagram is given in
Fig.~\ref{fig9rs}. There are two phases: a liquid phase above the
solid line. For $T> T_K$ it is a liquid where the replicas become
uncorrelated for $\varepsilon \rightarrow 0$ whereas for $T< T_K$
and $m<m^*(T)$ it is 
\index{molecular liquid}``molecular liquid'' 
where the
particles of the $m$ replica form ``molecules'' even for
$\varepsilon \rightarrow 0$. Below the solid line there is a glass
phase. Since $S_c(T)=0$ in the glass phase, Eq.~(\ref{eq92rs})
implies that $\phi(m,T)$ is \textit {independent} on $m$, i.e. it
is:
\begin{equation} \label{eq93rs}
\phi (m,T) = \phi(1,T)
\end{equation}
for all $m \geq m^*(T) $ and $T<T_K$. On the other hand $\phi
(m,T)$ is continuous at the liquid-glass phase boundary $m^*(T)$:
\begin{equation}\label{eq94rs}
\phi_{\textrm{liquid}}(m^*(T),T) = \phi
_{\textrm{glass}}(m^*(T),T)
\end{equation}
combining Eq.~(\ref{eq93rs}) and Eq.~(\ref{eq94rs}) one arrives at
the important result:
\begin{equation}\label{eq95rs}
\Phi(T) \equiv \phi(1,T) = \phi_{\textrm{liquid}}(m^*(T),T) \quad
.
\end{equation}
Due to Eq.~(\ref{eq95rs}) one can calculate the free energy
$\Phi(T)$ of the physical system from the free energy of the
replica system in its \textit{liquid} phase, despite $T<T_K$.
Since there are powerful techniques for the calculation of the
liquid free energy \cite{40rs} the relationship Eq.~(\ref{eq95rs})
allows to calculate $\Phi(T)$ from $\phi(m,T)$ and the latter also
allows to determine the configurational entropy $S_c(f,T)$ from
Eq.~(\ref{eq91rs}) and Eq.~(\ref{eq92rs}) by eliminating $m$.

\subsection{Microscopic description}
In the last section we have shown that the free energy $\Phi(T)$
and the configurational entropy $S_c(f,T)$ can be obtained from
$\Phi(m,T)$, the free energy of a replica system in its \textit {
liquid} phase. Now we will describe how this can be performed from
first principles.

The potential energy of a N-particle system in a volume $V$ may be
given by pair potentials $v(\vec{x})$:
\begin{equation} \label{eq96rs}
V(\vec{x}_1,\ldots,\vec{x}_N)= \sum \limits _{i<j} v(\vec{x}_i -
\vec{x}_j) \quad .
\end{equation}
Let us consider $m$ \textit {identical} systems
\index{replicas}(replicas) 
which are weakly coupled by an attractive pair
potential $w(\vec{x}_i^a - \vec{x}_j^b)$, which is considered to
be short ranged. Then the total potential energy is given by:
\begin{equation}\label{eq97rs}
V^\varepsilon _m (\{\vec{x}_i ^a\}) = \sum \limits _ {a=1}^m
V(\vec{x}_1^a, \ldots,\vec{x}_N^a)+\varepsilon\sum \limits _{a<b}
\sum \limits _{i<j} w(\vec{x}^a_i - \vec{x}_j^b)
\end{equation}
where $\vec{x}_i^a$ is the position of particle $i$ in replica $a$
and $\varepsilon \geq 0$ is an infinitesimal coupling constant.
Note, that $w$ breaks the \Index{replica permutational symmetry}.
It acts like a symmetry breaking magnetic field in case of a
ferromagnet. For $\varepsilon>0$ the thermodynamic limit $N
\rightarrow \infty$ forces the replicas into the same local
minima. Taking the limit $\varepsilon \rightarrow \infty$
afterwards leaves the replicas in the same state for $T< T_K$ and
makes them uncorrelated for $T>T_K$ (cf.~discussion in section
4.1). Therefore for $T<T_K$ and $\varepsilon>0$ it is (using an
appropriate labelling of the particles):
\begin{equation} \label{eq98rs}
\vec{x}_i^1 \approx \vec{x}_i^2\approx \cdots \approx \vec{x}_i^m
\end{equation}
Therefore we can introduce center of mass coordinates $\vec{X}_i$
of an ``$m$-atomic molecule'' and relative coordinates
$\vec{u}_i^a$ such that
\begin{equation} \label{eq99rs}
\vec{x}_i^a = \vec{X}_i+ \vec{u}_i ^a
\end{equation}
where
\begin{equation} \label{eq100rs}
\vec{X}_i = \frac 1 m \sum \limits _{a=1}^m \vec{x}_i^a
\end{equation}
and ${\vec{u}_i^a}$ has to fulfil the $N$ constraints:
\begin{equation}\label{eq101rs}
\sum \limits _{a=1}^m \vec{u}_i^a=0
\end{equation}
for all $i$.

\begin{figure}[t]
\unitlength1cm
\begin{picture}(11,7)
\centerline{\psfig{file=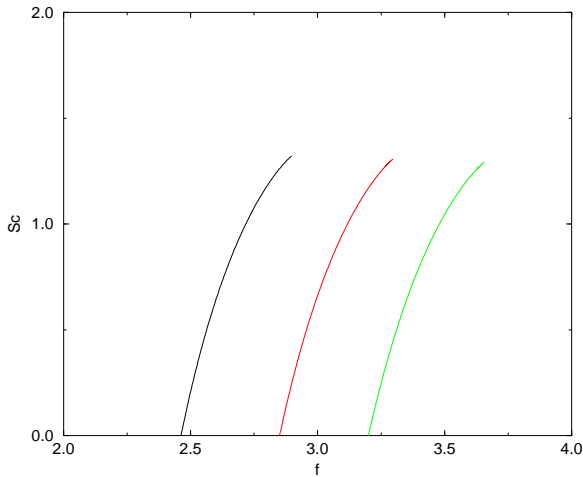,angle=-90,width=9cm}}
\end{picture}
\caption{ \label{fig10rs} Microscopic result for $S_c(f,T)$ for
$T=0.05,0.1,0.15$ (from left to right) for a system with soft
sphere potential (see also text).}
\end{figure}

The classical partition function (configurational part) is given
by:
\begin{equation}\label{eq102rs}
Z^\varepsilon _m (T,N)= \frac {1}{N!}\int \prod \limits _{i,a}
d^dx^a_i \exp[-\beta V^\varepsilon _m ( \{ \vec{x}_i^a \} )]
\end{equation}
where $d$ is the spatial dimension. Making use of
Eq.~(\ref{eq99rs}) this leads to
\begin{eqnarray}\label{eq103rs}
Z^\varepsilon _m (T,N)= \frac {1}{N!} \int \prod \limits _i d^d
X_i \int \prod \limits _ {i,a} d^d u^a_i \\ \nonumber \prod
\limits ^d _{i=1} \Big(m^d \delta \Big(\sum \limits _a
\vec{u}_i^a\Big)\Big) \exp [-\beta V^\varepsilon _m
(\{\vec{X}_i\}, \{\vec{u}_i^a\})]
\end{eqnarray}
where the $\delta$-function accounts for the constraints
Eq.~(\ref{eq101rs}). Due to Eq.~(\ref{eq98rs}), it is
$|\vec{u}_i^a|\ll 1$. Therefore a \Index{harmonic approximation}
can be applied which yields:
\begin{eqnarray} \label{eq104rs}
Z^\varepsilon _m (T,N)=Z _m (T,N) \cong \frac {1}{N!}\int \prod
\limits_i d^d X_i \int \prod \limits_{i,a} d^d u^a_i \cdot
\nonumber \\
\cdot \prod \limits_i \Big(m^d \delta \Big(\sum \limits_a
\vec{u}^a_i\Big)\Big)\cdot
\exp \Big\{ -\beta \Big[ m V ( \{ \vec{X}_i \} ) +  \nonumber \\
+ \frac{1}{2}  \sum \limits_a \sum \limits_{i\mu , j\nu} M_{i \mu,
j \nu} (\{\vec{X}_i \})(u^a_{i \mu} - u^a_{j\nu})^2 \Big] \Big\}
\end{eqnarray}
where $(M_{i \mu,j\nu})$ is the \Index{Hessian matrix} of V and
$\vec{u}^a_{i\mu}$ is the $\mu$-component of $\vec{u}^a_i$. Note
that we were allowed to put $\varepsilon = 0$, because we already
assumed that Eq.~(\ref{eq98rs}) holds. Using an integral
representation of the $\delta$-function the integral with respect
to $\vec{u}^a_{i\mu}$ are Gaussian and can be performed which
involves
\begin{equation}\label{eq105rs}
\Big[\prod \limits ^{Nd}_{k=1} \lambda_k( \{ \vec{X}_i \}
)\Big]^{1/2} \Big[\prod \limits ^{Nd}_{k=1} \lambda_k( \{
\vec{X}_i \} )\Big]^{-m/2} = \exp\Big[- \frac{m-1}{2} \textrm{Tr}
\ln (M_{i \mu , j\nu} ( \{ \vec{X}_i \}) )\Big]
\end{equation}
where $\lambda _k$ are the eigenvalues of $(M_{i \mu , j\nu})$.
The result can be brought into the following form:
\begin{equation} \label{eq106rs}
Z _m (T,N) \cong (c_m(T))^N Z_1(T/m,N) \Big\langle
\exp\Big[-\frac{m-1}{2} \textrm{Tr}\ln (M_{i \mu, j\nu}
(\{\vec{X}_i\}))\Big] \Big\rangle (T/m)
\end{equation}
where:
\begin{equation}\label{eq107rs}
c_m(T)= m ^{\frac {d} {2}}(2 \pi T)^{\frac {d} {2}(m-1)} \quad .
\end{equation}
$Z_1(T/m,N)$ is the partition function of the original system, but
at a temperature $T/m $ and $\langle ( \cdot )\rangle$ denotes
averaging with respect to $\exp[-\beta m V (\{\vec{X}_i\})]/Z_1 \\
(T/m,N)$. In a next step one uses the approximation:
\begin{equation}\label{eq108rs}
\Big\langle \exp \Big[ \cdots\Big] \Big\rangle \approx \exp
\Big[-\frac{m-1}{2} \langle \textrm{Tr} \ln (M_{i \mu, j\nu} ( \{
\vec{X}_i \} ))\rangle \Big] \quad .
\end{equation}
The r.h.s. of Eq.~(\ref{eq108rs}) could be calculated if the
density distribution of the eigenvalues would be known, which is
not the case. Therefore, some more approximations are necessary.
We will not present these technical manipulations which can be
found in Ref.~\cite{50rs} but present the idea. Let us skip for a
moment the logarithm in Eq.~(\ref{eq108rs}). Then we have to
calculate $\langle \textrm{Tr}(M_{i \mu, j\nu} ( \{ \vec{X}_i \}
))\rangle = \sum \limits _{i,\mu} \langle M_{i \mu, i\mu} ( \{
\vec{X}_i \} )\rangle $. Since $M_{i \mu, i\mu} ( \{ \vec{X}_i \}
) = \sum \limits _j (\partial ^2v/\partial r^2_\mu) ( \vec{r}=
\vec{X}_i - \vec{X}_j)$ this involves the average of the second
derivatives of the pair potential $v(\vec{r})$. Assuming that
$M_{i \mu, i\mu}$ do not fluctuate much one obtains finally

\begin{equation}\label{eq109rs}
\langle \textrm{Tr} (M_{i \mu, j\nu} ( \{ \vec{X}_i \} ))\rangle
\Big(\frac T m\Big) \approx \frac N d \int d^drg\Big(\vec{r},\frac
T m\Big) \vec{\nabla}^2v(\vec{r})
\end{equation}
i.e. $\langle \textrm{Tr} (M_{i \mu, j\nu} ( \{ \vec{X}_i \}
))]\rangle (T/m)$ can be related to the \Index{pair distribution
function} at $T/m$. In a similar way one can express $\langle
\textrm{Tr}\ln (M_{i \mu, j\nu} ( \{ \vec{X}_i \} ))]\rangle
(T/m)$ by $g(\vec{r},T/m)$:
\begin{equation} \label{eq110rs}
\langle \textrm{Tr} \ln (M_{i \mu, j\nu} ( \{ \vec{X}_i \}
))]\rangle \Big(\frac T m \Big) \approx N {\cal G} \Big[g(\vec{r};
\frac T m)\Big] \; .
\end{equation}
Using these approximations we end up with the \Index{free energy}
$\Phi (m,T) = - T \lim \limits _ {N \rightarrow \infty} N ^{-1}
\ln Z_m(T,N)$:
\begin{equation}\label{eq111rs}
\beta \Phi (m,T) \approx - \ln c_m(T) + \Phi (T/m) + \frac 1 2
(m-1){\mathcal{G}}[g(\vec{r}; T/m)] \quad .
\end{equation}
This result demonstrates what has been discussed in section 4.1.
The free energy $\Phi (m,T)$ of the replica system is given by the
free energy of the original system $(m=1)$ at $T/m$ and by a
functional ${\mathcal{G}}$ of the pair distribution function of
the original system, also at $T/m$. In the liquid phase $\Phi
(T/m)$ can also be expressed by $g(\vec{r},T/m)$ \cite{40rs}.
Therefore, up to an irrelevant term $\ln c_m (T)$ one has
succeeded to express $\Phi (m,T)$ by the pair distribution
function at a temperature $T/m$. If $T$ is below $T_K$, one has to
choose $m$ small enough in order to be in the liquid phase. As
already mentioned above there are powerful methods \cite{40rs} to
calculate $g(\vec{r},T/m)$ for the liquid phase. Finally, the such
obtained $\Phi(m,T)$ allows to determine $\Phi(T)$ and $S_c(f,T)$
as described in section 4.1. In Figure \ref{fig10rs} we show
$S_c(f,T)$ obtained by Mézard and Parisi \cite{49rs} for a
three-dimensional system $(d=3)$ with density one and a soft
sphere pair potential $v(\vec{r})= |\vec{r}|^{-12}$. The
$f-$dependence has the qualitatively correct behavior, on which
the discussion in section 4.1 has been founded. Using the
condition Eq.~(\ref{eq86rs}) for the static glass transition point
with $T_s \equiv T_K$ these authors obtained $T_K \cong 0.194$ or
$\Gamma _K \cong 1.51$ for the dimensionless coupling parameter
$\Gamma  = nT^{-1/4}$. This value is in a reasonable range and is
close to the values $\Gamma _K ^{sim}\cong 1.60$ \cite{52rs} and
1.46 \cite{53rs} from a MD-simulation of \textit{binary} soft
sphere liquids. Since the authors of Ref.~\cite{53rs} claim that
their results are compatible with MCT, it is not quite obvious
whether the numerical values yield $\Gamma _c  = nT_c^{-1/4}$ or
not.

\section{SUMMARY}
In this article we have reviewed phenomenological and microscopic
theories for the structural glass transition. The phenomenological
approaches rely on several assumptions which are not proven to be
correct. Although they are connected with appealing ``physical
pictures'' their predictive power is limited. However, one class
of them has obtained a microscopic justification by the replica
theory for structural glasses \cite{45rs,49rs,50rs}. This theory,
based on first principles, predicts a \textit {static} glass
transition at the \Index{Kauzmann temperature} $T_K$ where the
\Index{configurational entropy} $S_c(T)$ vanishes, as stated in
the Adams-Gibbs-theory (1st chapter). Although one can avoid the
use of \Index{replicas} \cite{54rs}, the replica theory has a
certain beauty because the several equivalent glass phases with
same  free energy can be described by copying the system $m$ times
and introducing a weak coupling between the copies (replicas).
This coupling acts as a 
\index{symmetry breaking field}``symmetry'' breaking field
similar to a magnetic field for a ferromagnet. Analytically
continuing $m$ to positive real numbers allows to relate the
thermodynamical properties of the glass phase, i.e. for $T<T_K$,
to those of the liquid phase, provided $m$ is taken small enough
with respect to one.

Quite a different type of transition is obtained from \Index{mode
coupling theory} \cite{27rs,28rs,29rs}. MCT is a dynamical theory,
in contrast to replica theory. It provides an equation of motion
for, e.g. the spatial Fourier transform $\Phi (\vec{q},t)$ of the
normalized density correlator of a simple liquid. Extension to
molecular liquids is straightforward
\cite{15rs,30rs,31rs,32rs,33rs}. MCT predicts the existence of a
\textit {dynamical} glass transition at a temperature $T_c$ where
the dynamics changes qualitatively. Above $T_c$ the correlator
$\Phi$ decays to zero, and converges to the \Index{nonergodicity
parameter} $f(\vec{q})>0$, below $T_c$. The nonergodicity
parameters vary \textit {discontinuously} at $T_c$ and can be
interpreted as glass order parameters. Besides this, MCT makes
several new predictions. Close to $T_c$ there exist two scaling
laws, the \textit {critical} and the \textit {von Schweidler law}.
The corresponding time scales $t_\sigma (T)$ and $\tau (T)$
exhibit power law divergence at $T_c$. For times much larger than
a typical microscopic time $t_0$ and much smaller than the
$\alpha$-relaxation time scale $\tau (T)$ the $\vec{q}$- and
$t$-dependence of $\Phi (\vec{q},t)$ factorize, which is a very
strong statement. Furthermore, all the exponents of those power
laws can be obtained from \textit {one} parameter $\lambda$, only,
the \Index{exponent parameter}. $\lambda$ is determined by the
static correlator $S(\vec{q})$ at $T_c$. This proves the very
microscopic nature of MCT. Although $\lambda$ and therefore the
exponents are dependent on the physical system, i.e. on the
interaction, they are \textit{universal} for \textit{all}
correlators of one system which couple to the density fluctuations
\cite{27rs}.

These two microscopic theories in some sense are complementary to
each other and yield two different glass transition points. It is
interesting that the existence of a static and a dynamical glass
transition had also been found for mean field spin glasses with
discontinuous order parameter. There it has even been speculated
that \Index{spin glasses} are quite similar to \Index{structural
glasses} \cite{8rs}. For the mean field spin glasses both
transitions are related to \textit {singularities} and occur at
$T_K$ and $T_c$ with $T_K < T_c$. Therefore both transitions are
\textit{sharp}. Concerning MCT (as described in the 3rd chapter),
sometimes called \Index{idealized mode coupling theory}, it has
been shown \cite{35rs,55rs,56rs} that the singularity is removed,
due to \Index{ergodicity restoring processes}. Nevertheless, if
the time scale of these processes is much larger than $\tau (T)$,
one can observe in a certain time and temperature window the
dynamical behavior as predicted from the idealized theory
\cite{27rs,28rs,29rs}. Deviations start to emerge very close to
$T_c$. This fact has some importance for the static glass
transition. For mean field spin glasses it was proven
\cite{57rs,58rs} that the system does not relax to equilibrium
below $T_c$. Therefore the static transition at $T_K$ is masked.
Due to the ergodicity restoring processes this is not anymore true
for systems with finite range interactions. Despite of that, it is
by far not clear that there is a minimum free energy
$f_{\textrm{min}}(T)$ at which the configurational entropy
$S_c(f,T)$ vanishes linearly with a slope $s_0(T)=(\partial
S_c/\partial f) (f_{\textrm{min}}(T),T) < \infty$. This is a very
strong assumption which may not be fulfilled in general. For real
systems, i.e.~in finite dimensions, with short range interactions
the physical picture described in section 4.1 probably does not
hold, i.e.~ there are not an exponential number of states with
density $\rho^{(\alpha)} (\bf x)$ and infinite life time. However,
there might exist systems with large enough frustration being
close to the idealized situation, at least on a finite time scale.
The same holds for MCT. For instance, the ergodicity restoring
processes seem to be extremely weak for \Index{colloidal
systems}. Indeed it has been shown that the dynamics of colloids
can be described over many decades in time by MCT \cite{59rs}.

Independent on whether the singularities of both microscopic
theories exist or not, the progress which has been made is
significant, this is particularly true for MCT. The number of
experiments and simulations \cite{12rs,13rs,14rs,15rs} (see also
the contribution by U. Buchenau in this monograph) having found
consistency with the MCT-predictions in an appropriately chosen
time and temperature interval is enormous, despite the deviations
very close to $T_c$. One may hope that the few tests \cite{60rs}
of replica theory may be continued in order to check its validity
in more detail. And an extension including time-dependence would
be desirable as well.

Both theories probably describe idealized situations only. As
mentioned above MCT has been extended \cite{55rs,56rs} to include
ergodicity restoring processes. But this extended version has not
really been tested, because it is rather involved. Describing the
ergodicity restoring processes by a single parameter $\delta$
there was a comparison with experimental data which is
satisfactory \cite{61rs}. Whether \Index{replica theory} can be
extended in case that the singularity is spurious is not clear.
Such an extension might also lead to a much more complicated
mathematical structure. In that case it might be much better to
restrict to the idealized version, which holds for both theories.

Nevertheless there are still many challenging problems left. Let
us mention some of them:

\begin{itemize}
\item[{(i)}]
\hspace{0,01cm}further investigations of similarities and
dissimilarities between systems with and without quenched disorder

\item[{(ii)}]
\hspace{0,01cm} glass transition in \Index{lattice gas models}
\cite{62rs}

\item[{(iii)}]
\hspace{0,01cm} connection between the 
\index{potential energy landscape}``potential energy landscape'' 
properties \cite{16rs,17rs,18rs} with MCT and replica
theory; e.g. it has been shown that the \Index{saddle index}
vanishes at the MCT-temperature $T_c$ \cite{63rs}

\item[{(iv)}]
\hspace{0,01cm} investigation of models with trivial statics which
may not exhibit a static glass transition but a MCT-like one
\cite{64rs}

\item[{(v)}]
\hspace{0,01cm} investigations of models without or weak static
correlations \cite{65rs}, where replica theory and MCT (in its
present form) do not yield a glass transition, although found in
simulations \cite{66rs}

\item[{(vi)}]
\hspace{0,01cm} \Index{nonequilibrium} behavior (\Index{aging})
\end{itemize}

\bigskip
\bigskip

{\bf ACKNOWLEDGMENT}: I gratefully acknowledge valuable comments
on this manuscript by Pablo G. Debenedetti, Wolfgang G\"otze and
Marc M\'ezard. Figures~\ref{fig9rs} and ~\ref{fig10rs} were
provided by Marc M\'ezard and Figures~\ref{fig1rs} and
~\ref{fig2rs} were produced by Michael Ricker. I also would like
to thank both for their support.

\clearpage
%
%
%
%

%
%


\printindex
\end{document}